\documentclass[%
 reprint,
superscriptaddress,
 amsmath,amssymb,
 aps,
pra,
]{revtex4-2}

\usepackage{graphicx}
\usepackage{dcolumn}
\usepackage{bm}

\newcommand{\ket}[1]{|#1\rangle}

\newcommand{\bra}[1]{\langle #1|}

\DeclareMathAlphabet{\pazocal}{OMS}{zplm}{m}{n}
\DeclareMathOperator{\tr}{tr}
\usepackage[colorlinks=true, linktocpage=true]{hyperref}
\usepackage{blindtext}
\hypersetup{colorlinks=true, linktoc=all, citecolor=blue,linkcolor=blue,urlcolor=blue}

\usepackage{orcidlink}

\begin{document}

\preprint{APS/123-QED}

\title{Microscopic derivation of a completely positive master equation for the description of Open Quantum Brownian Motion of a particle in a potential}

\author{Ayanda Zungu\orcidlink{0000-0001-5972-9179}}
\email[]{ayanda.zungu@nwu.ac.za}
\affiliation{Discipline of Physics, School of Agriculture and Science, University of KwaZulu-Natal, Durban 4001, South Africa}
\affiliation{Centre for Space Research, North-West University, Mahikeng 2745, South Africa}

\author{Ilya Sinayskiy\orcidlink{0000-0002-3040-0051
}}
\email[]{sinayskiy@ukzn.ac.za}
\affiliation{Discipline of Physics, School of Agriculture and Science, University of KwaZulu-Natal, Durban 4001, South Africa}

\affiliation{National Institute for Theoretical and Computational Sciences (NITheCS), Stellenbosch, South Africa}

\author{Francesco Petruccione\orcidlink{0000-0002-8604-0913
}}
\email[]{sinayskiy@ukzn.ac.za}
\affiliation{National Institute for Theoretical and Computational Sciences (NITheCS), Stellenbosch, South Africa}

\affiliation{School of Data Science and Computational Thinking and Department of Physics, Stellenbosch University, Stellenbosch 7604, South Africa}
 
\date{\today}

\begin{abstract}
Open Quantum Brownian Motion (OQBM) was introduced as a scaling limit of discrete-time open quantum walks.
This limit defines a new class of quantum Brownian motion, which incorporates both the external and internal degrees of freedom of the Brownian particle. We consider a weakly driven Brownian particle confined in a harmonic potential and dissipatively coupled to a thermal bath. Applying the rotating wave approximation (RWA)
to the system-bath interaction Hamiltonian, we derive a completely positive Born-Markov master equation for the reduced dynamics. We express the resulting master equation in the coordinate representation and, utilizing the adiabatic elimination of fast variables, derive a completely positive hybrid quantum-classical master equation that defines OQBM. We illustrate the resulting dynamics using examples of initial Gaussian and non-Gaussian distributions of the OQBM walker. Both examples reveal the emergence of Gaussian distributions in the limiting behavior of the OQBM dynamics, which closely matches that of the standard OQBM. With the help of the obtained OQBM master equation, we derive the equations for the $n$-th moments and the cumulants of the position distribution of the open Brownian walker.
We subsequently solve these equations numerically for Gaussian initial distributions across various parameter regimes. Notably, we find that the third-order cumulant is nonzero, indicating that the dynamics' intrinsic generator is non-Gaussian.
\end{abstract}

\maketitle

\section{Introduction}

Realistic quantum systems are not isolated and interact with their environment (or a ``bath'') to some extent~\cite{breuer2002theory}. These interactions typically introduce disruptive noise, leading to quantum dissipation and decoherence, which constitute the primary obstacle to realizing quantum computers and other quantum devices~\cite{schlosshauer2007decoherence,gardiner2004quantum}. The theory of open quantum systems provides a framework for understanding fundamental problems in quantum mechanics, such as the problem of the quantum-to-classical transition through the bath-induced decoherence mechanism~\cite{bai2017classical,kovacs2017quantum}.

A primary tool for investigating the dynamics of open systems is the family of Born-type master equations (2nd order approximation in the system-bath interaction), which describes the evolution of the reduced density matrix of the system of interest obtained by taking the partial trace over the degrees of freedom of the thermal bath~\cite{breuer2002theory}. One of them is the celebrated master equation in the  
Gorini-Kossakowski-Sudarshan-Lindblad (GKSL) form, which guarantees complete positivity of the reduced density matrix (often called the Lindblad master equation) ~\cite{lindblad1976generators,gorini1976completely, Darek2017GKSL}.
The Lindblad master equation ensures that the time evolution of the system of interest maintains the critical properties of density matrices: Hermiticity, unit trace, and positive semidefiniteness throughout the evolution. 
The stronger condition of complete positivity is required to ensure physically consistent dynamics, particularly for composite systems~\cite{nielsen2010quantum}.

The ubiquitous model of an open quantum system is the quantum Brownian motion (QBM), particularly the QBM described by the Caldeira-Leggett (CL) model~\cite{caldeira1983path, CALDEIRA1983374}. This model is one of several models representing a Brownian particle moving in a quadratic potential and weakly interacting with a thermal bath modelled by an infinite set of independent harmonic oscillators, and is valid at relatively high temperatures, specifically for Markovian coupling to the thermal bath. 
It has served as a foundational model for numerous studies, being both rich in physical insight and largely analytically tractable.
Despite its simplicity, the CL model has become widely used in condensed matter physics~\cite{raju2016quantum,yabu1998dissipative,marquardt2004relaxation} due to its broad applicability and ability to describe dissipation and decoherence in quantum systems. 
However, the Born-Markov master equation derived from the CL model does not fall within the Lindblad class~\cite{lindblad1976generators,gorini1976completely}, which means that the resulting semigroup is not completely positive~\cite{L.Diósi_1993,homa2019positivity}, and many efforts have been made in the literature to resolve this issue~\cite{L.Diósi_1993,diosi1993calderia,vacchini2000completely}. 

This paper focuses on \textit{Open Quantum Brownian Motion} (OQBM)~\cite{bauer2013open,bauer2014open}, a new type of quantum Brownian motion where the probability of finding the walker at a particular position depends not only on the interaction with the bosonic bath but also on the state of the internal degree of freedom of the Brownian particle. Bauer~\textit{et al.}~\cite{bauer2013open,bauer2014open} introduced these walks as a scaling limit to discrete-time open quantum walks (OQWs)~\cite{ATTAL20121545,attal2012open, Sinayskiy_2012} at the level of the Kraus maps without any mention of future physical realizations.
Recently, a microscopic derivation of OQBM for the case of a free Brownian particle and decoherent interaction with the thermal bath has been developed~\cite{sinayskiy2015microscopicbrown,sinayskiy2017steady} and analyzed ~\cite{Carlos2025}. 

In our previous work~\cite{zungu2025adiabatic}, we used the adiabatic elimination of fast variables ~\cite{smoluchowski1916brownsche,van1985elimination,gardiner1985handbook,kramers1940brownian} to derive OQBM for a generic dissipative scenario. Even though this model~\cite{zungu2025adiabatic} led to the correct OQBM dynamics~\cite{bauer2013open,bauer2014open,sinayskiy2015microscopicbrown,sinayskiy2017steady}, we encountered nonphysical results for some initial conditions and evolution times due to the limitations of the standard CL model~\cite{caldeira1983path, CALDEIRA1983374}.
In this paper, as an extension of the previous study~\cite{zungu2025adiabatic}, we now derive a completely positive master equation for OQBM in the case of a weakly driven open Brownian particle confined within a harmonic potential and dissipatively coupled to a thermal bath. 

To derive the reduced density matrix, we employ the traditional techniques of the theory of open quantum systems~\cite{breuer2002theory}, where one starts from the microscopic Hamiltonian comprising the Hamiltonian describing a weakly driven open Brownian particle, the Hamiltonian of the thermal bath, and the system-bath interaction Hamiltonian. As a first step in the microscopic derivation, we apply the Born-Markov approximation and trace out the bath degrees of
freedom~\cite{breuer2002theory,schlosshauer2007decoherence}. By performing the \textit{rotating wave approximation} (RWA)~\cite{thimmel1999rotating} to the system-bath interaction Hamiltonian, we derive a completely positive Born-Markov master equation for the reduced density matrix.

To demonstrate that the resulting master equation can be written as OQBM~\cite{bauer2013open,bauer2014open}, we need to show that it can be represented in a diagonal representation in position, and we achieve this by starting from a generic non-diagonal representation. We then perform the adiabatic elimination of fast variables~\cite{smoluchowski1916brownsche,van1985elimination,gardiner1985handbook,kramers1940brownian, Gardiner1984} in the coordinate representation and obtain a master equation that defines OQBM~\cite{bauer2013open,bauer2014open,sinayskiy2015microscopicbrown,sinayskiy2017steady,zungu2025adiabatic}. 
The adiabatic elimination of fast variables is similar to the strong-friction limit of the Quantum Smoluchowski equation~\cite{Ankerhold2001}. The derived completely positive OQBM master equation contains diffusive, dissipative, and ``decision-making'' terms, and it has the same structure as the master equation proposed by Bauer~\textit{et al.}~\cite{bauer2013open,bauer2014open} and demonstrated by~\cite{sinayskiy2015microscopicbrown,sinayskiy2017steady}. We also recently derived this type of master equation~\cite{zungu2025adiabatic}.
The master equation governing OQBM is a notable example of a hybrid quantum-classical master equation~\cite {PhysRevA.107.062206, Diósi_2014}, a class of dynamical equations that has attracted growing interest in recent literature~\cite{layton2024healthier,layton2024classical,oppenheim2023postquantum,oppenheim2022two,halliwell1998effective,tilloy2024general}.

The paper has the following structure. In Sec.~\ref{microscopic_deriv}, we formulate the microscopic model describing a weakly driven open Brownian particle and perform the microscopic derivation of a completely positive Born-Markov master equation for the reduced dynamics. In Sec.~\ref{elimination}, we present the adiabatic elimination of fast variables method and derive a completely positive master equation that defines OQBM. In Sec.~\ref{numerical_examples}, we demonstrate the derivation by showing examples of OQBM dynamics for Gaussian and non-Gaussian initial distributions of the open Brownian particle. In Sec.~\ref{moments_cumulants}, we use the obtained OQBM master equation to derive equations for the $n$-th moments and the cumulants of the position distribution of the OQBM walker, respectively. These equations are solved numerically for Gaussian initial distributions across various parameter regimes. Finally, in Sec.~\ref{conclussion}, we conclude.

\section{Microscopic derivation of completely positive Born-Markov master equation}\label{microscopic_deriv}

The OQBM model consists of an open Brownian particle of mass $m$, with position operator $\hat{x}$, representing the external degree of freedom coupled linearly to an infinite set of harmonic oscillators constituting the thermal bath. The quantum internal degree of freedom is described by a two-level system (2LS). From a microscopic perspective, the total Hamiltonian of the combined system and bath is given by
\begin{equation}
    \hat{H} = \hat{H}_S+\hat{H}_B+\hat{H}_{SB},
\end{equation}
\noindent
where $\hat{H}_S$, $\hat{H}_B$, and $\hat{H}_{SB}$ denote the Hamiltonians of the system (open Brownian particle), the bath, and the system-bath interaction Hamiltonian, respectively. 
The system Hamiltonian $\hat{H}_S$ is given by
\begin{align}\label{sysHami}
\hat{H}_S = \frac{\hat{p}^2}{2m}+\frac{m\omega^2\hat{x}^2}{2}+\frac{\hbar\omega_0}{2}\hat{\sigma}_z+\hbar\Omega\hat{\sigma}_x,
\end{align}
\noindent
where $\frac{\hat{p}^2}{2m}$ is the Hamiltonian of the Brownian particle, $\frac{m\omega^2\hat{x}^2}{2}$ is the quadratic potential trapping the particle, $\frac{\hbar\omega_0}{2} \hat{\sigma}_z$ is the Hamiltonian of the 2LS with $\omega_0$ denoting its transition frequency.
The last term $\hbar\Omega\hat{\sigma}_x$ describes a weak driving of the inner degree of freedom ($\Omega\ll\omega_0$)~\cite{carmichael1999statistical,scully1999quantum}. The operators $\hat{\sigma}_{i=x,y,z}$ are the standard Pauli matrices, and the position operator $\hat{x}$ together with the momentum operator $\hat{p}$  in the system~(\ref{sysHami}) satisfy the fundamental commutation relations $[\hat{x},\hat{p}]=i\hbar$. 
The thermal bath $\hat{H}_B$ is given by
\begin{align}
\hat{H}_B = \sum_n \hbar \omega_n \hat{a}_n^\dagger\hat{a}_n,
\end{align}
\noindent
where $\omega_n$ are the natural frequencies of the associated oscillators and the bosonic operators $\hat{a}_n^\dagger$ and $\hat{a}_n$ denote the creation and destruction operators, respectively, and together, they satisfy the standard commutation relations $[\hat{a}_n,\hat{a}_m^\dagger]=\delta_{n,m}$. The system-bath interaction Hamiltonian $\hat{H}_{SB}$, describing the coupling between both degrees of freedom of the Brownian walker and the bath, is given by
\begin{align}\label{h_int}
\hat{H}_{SB} = \hbar\bigl(\hat{x}+\alpha\hat{\sigma}_x\bigl)\sum_n g_n \bigl(\hat{a}_n+\hat{a}_n^\dagger\bigl),
\end{align}
\noindent
where $g_n$ represents the coupling strength between the thermal bath and the Brownian particle, and a constant $\alpha$ is a relative coupling strength. 

To derive the quantum master equation for the reduced dynamics $\hat{\rho}_S(t)$, we assume that the system is weakly coupled to the thermal bath, allowing us to apply the standard Born-Markov approximation~\cite{breuer2002theory,schlosshauer2007decoherence}.
Under these assumptions, the reduced density matrix evolves as,
\begin{align}\label{generic_master}
    \frac{d}{dt}\hat{\rho}_S(t) &= -\frac{1}{\hbar^2}\int_0^\infty d\tau \tr_B\Bigl[\hat{H}_{SB}(t),
    \bigl[\hat{H}_{SB}(t-\tau),\nonumber\\
    &\hat{\rho}_S(t)\otimes\hat{\rho}_B\bigl]\Bigl],
\end{align}
\noindent
where $[\cdot,\cdot]$ stands for the commutator, $\hat{H}_{SB}(t)$ describes the system-bath interaction Hamiltonian in the interaction picture and $\hat{\rho}_B$ denotes the density matrix of the bosonic bath. The bath is assumed to be in thermal equilibrium at temperature $T=(k_B \beta)^{-1}$, i.e., $\hat{\rho}_B=\mathcal{Z}^{-1} \exp[-\beta \hat{H}_B]$, where  $k_B$ is the Boltzmann constant and the partition function is $\mathcal{Z}=\tr_B[\exp(-\beta\hat{H}_B)]$. 

However, Eq.~(\ref{generic_master}) does not guarantee that the resulting dynamical equation will take the GKSL form~\cite{breuer2002theory,gorini1976completely,lindblad1976generators}. To derive the master equation in the GKSL form, one needs to apply the RWA~\cite{thimmel1999rotating} directly to the interaction Hamiltonian~(\ref{h_int}). 
By doing so, the system-bath interaction Hamiltonian~(\ref{h_int}) takes now the following form:
\begin{align}\label{h_int_rwa}
\hat{H}_{SB} &= \hbar x_0\sum_n g_n (\hat{a}^\dagger\hat{a}_n+\hat{a}\hat{a}_n^\dagger)\nonumber\\
&+\hbar \alpha\sum_n g_n \bigl(\hat{a}_n\hat{\sigma}_++\hat{a}_n^\dagger\hat{\sigma}_-\bigl),
\end{align}
\noindent
where $x_0 = \sqrt{\hbar/2m\omega}$, and $\hat{\sigma}_\pm$ are the Pauli raising and lowering operators, satisfying the commutation relation $[\hat{\sigma}_+,\hat{\sigma}_-]=\hat{\sigma}_z$.
The general Born-Markov master equation~(\ref{generic_master}), when applied to our
system, takes the form:
\begin{align}
    \frac{d}{dt} \hat{\rho}_S (t) =& -\frac{i}{\hbar}\bigl[\hat{H}_S,\hat{\rho}_S\bigl]-\frac{1}{\hbar^2}\int_0^\infty d\tau \tr_B\Bigl[\hat{H}_{SB}(0),\nonumber\\
    &\bigl[\hat{H}_{SB}(-\tau),\hat{\rho}_S(t)\otimes\hat{\rho}_B\bigl]\Bigl].
\end{align} 
The system-bath interaction Hamiltonian $\hat{H}_{SB}(-\tau)$ in the interaction picture can be written as
\begin{align}\label{interHam}
    \hat{H}_{SB}(-\tau) &= x_0\sum_n g_n\hat{a}^\dagger\hat{a}_ne^{i(\omega_n-\omega)\tau}+\mathrm{h.c.}\nonumber\\
    &+\alpha\sum_n g_n\hat{a}_n\hat{\sigma}_+e^{i(\omega_n-\omega_0)\tau}+\mathrm{h.c.}
\end{align}
\noindent
We assume that the external driving is weak ($\Omega\ll \omega_0$). This allows us to neglect its corresponding contributions when deriving the explicit form of $\hat{H}_{SB}(-\tau)$ in Eq.~(\ref{interHam}).

Using the explicit form of the system-bath interaction Hamiltonian in the interaction picture~(\ref{interHam}), applying the additional RWA, and choosing the bath spectral density $J(\tilde{\omega})$ such that $\sum_n |g_n|^2 \rightarrow \int d\tilde{\omega} J(\tilde{\omega})$, we derive the following equation for the reduced dynamics
\begin{align}\label{master_eq}
    \frac{d}{d t} \hat{\rho}_S(t) = \pazocal{L}_\mathrm{QHO}\hat{\rho}_S+\pazocal{L}_\mathrm{2LS}\hat{\rho}_S+\pazocal{L}_\mathrm{cross}\hat{\rho}_S,
\end{align}
\noindent
where the first term $\pazocal{L}_\mathrm{QHO}\hat{\rho}_S$ represents the dissipator of the quantum harmonic oscillator, which reads
\begin{align}\label{qho}
&\pazocal{L}_\mathrm{QHO}\hat{\rho}_S(t) = -\frac{i}{\hbar} \bigl[\hat{H}_\mathrm{QHO},\hat{\rho}_S\bigl]-\bar{\alpha}_1\bigl[\hat{x},[\hat{x},\hat{\rho}_S]\bigl]\nonumber\\
&-\bar{\alpha}_2\bigl[\hat{p},[\hat{p},\hat{\rho}_S]\bigl]+i\bar{\alpha}_3\Bigl(\bigl[\hat{p},\{\hat{x},\hat{\rho}_S\}\bigl]-\bigl[\hat{x},\{\hat{p},\hat{\rho}_S\}\bigl]\Bigl).
\end{align}
 \noindent
Above, $\{\cdot,\cdot\}$ denote the anti-commutator. The Hamiltonian describing the quantum harmonic oscillator $\hat{H}_\mathrm{QHO}$ and the constants appearing in Eq.~(\ref{qho}) are 
 \begin{align}
    & \hat{H}_\mathrm{QHO} = \frac{\hat{p}^2}{2m}+\frac{m\omega^2\hat{x}^2}{2}, \hspace{2mm} \bar{\alpha}_1=\frac{\pi }{4}J(\omega)(2n(\omega)+1),\nonumber\\
     &\bar{\alpha}_2=\frac{\pi J(\omega)}{4(m\omega)^2}(2n(\omega)+1),\hspace{2mm} \bar{\alpha}_3=\frac{\pi J(\omega) }{4m\omega},
 \end{align}
 \noindent
where $n(\omega)$ is the mean occupation number of the oscillator at frequency $\omega$. Equation~(\ref{qho}) represents a generalized CL-type~\cite{caldeira1983path,CALDEIRA1983374}  master equation with an extra term $\bigl[\hat{p},[\hat{p},\hat{\rho}_S]\bigl]$, which has been previously added in the literature to correct for the positivity violation constraint of the reduced density matrix~\cite{breuer2002theory,L.Diósi_1993,diosi1993calderia,vacchini2000completely}. The presence of $\bigl[\hat{p},[\hat{p},\hat{\rho}_S]\bigl]$ enables Eq.~(\ref{qho}) to be written in the GKSL form~\cite{breuer2002theory,gorini1976completely,lindblad1976generators} and just like in Eq.~(3.63) of Breuer and Petruccione~\cite{breuer2002theory}, one can choose the Lindblad operators, i.e., $\hat{F}_1=\hat{x}$, and $\hat{F}_2=\hat{p}$, and show that the Kossakowski matrix is given by
\begin{align}
[a_{ij}] = \begin{pmatrix}
        2\bar{\alpha}_1 &2i\bar{\alpha}_3\\
        -2i\bar{\alpha}_3 &2\bar{\alpha}_2\\
    \end{pmatrix}.
\end{align}
\noindent
From the above, it is straightforward to show that Eq.~(\ref{qho}) satisfies the positivity constraint, which reads
\begin{align}\label{kossa}
    \det[a_{ij}] &= 4\bar{\alpha}_1\bar{\alpha}_2-4\bar{\alpha}_3^2\nonumber\\
    &=\biggl(\frac{\pi J(\omega)}{m\omega}\biggl)^2 n(\omega)\Bigl(n(\omega)+1\Bigl)\ge 0.
\end{align}
\noindent
From the determinant of the Kossakowski matrix~(\ref{kossa}), it is clear that the derived master equation~(\ref{qho}) is completely positive for any bath temperature $T\ge 0$, including the zero-temperature limit, ensuring physically consistent dynamics at all temperatures.

The second term in Eq.~(\ref{master_eq}), $\pazocal{L}_\mathrm{2LS}\hat{\rho}_S$, has the form of the famous quantum optical master equation for a 2LS~\cite{carmichael1999statistical,scully1999quantum}, and represents the dissipation of a weakly driven internal degree freedom, which reads
\begin{align}\label{2ls}
\pazocal{L}_\mathrm{2LS}\hat{\rho}_S(t) &= -i\bigl[(\omega_0/2)\hat{\sigma}_z+\Omega\hat{\sigma}_x,\hat{\rho}_S\bigl]+i\beta_3\bigl[\hat{\sigma}_z,\hat{\rho}_S\bigl]\nonumber\\
    &+\beta_1\pazocal{L}[\hat{\sigma}_-,\hat{\sigma}_+]\hat{\rho}_S+\beta_2\pazocal{L}[\hat{\sigma}_+,\hat{\sigma}_-]\hat{\rho}_S,
\end{align}
\noindent
where,
\begin{align}
    &\beta_1 = 2\alpha^2\pi J(\omega_0)(n(\omega_0)+1), \hspace{2mm} \beta_2 = 2\alpha^2\pi J(\omega_0)n(\omega_0),\nonumber\\
     &\beta_3= \alpha^2\  \mathrm{P}\int d\tilde{\omega}\frac{J(\tilde{\omega})(n(\tilde{\omega})+1)}{\tilde{\omega}-\omega_0},
\end{align}
\noindent
and $\pazocal{L}[\hat{n},\hat{n}^\dagger]\hat{\rho}_S=\hat{n}\hat{\rho}_S\hat{n}^\dagger-(1/2)\{\hat{n}^\dagger\hat{n},\hat{\rho}_S\}$ denotes the dissipative super-operator in the GKSL form~\cite{breuer2002theory,lindblad1976generators,gorini1976completely}.
The last term in Eq.~(\ref{master_eq}), $\pazocal{L}_\mathrm{cross}\hat{\rho}_S$, represents a ``cross-term'' dissipator that accounts for dissipation arising from the interaction between external and internal degrees of freedom and reads
\begin{align}\label{cross}
&\pazocal{L}_\mathrm{cross}\hat{\rho}_S(t)=-\bar{\Omega}(\alpha_2'+\alpha_1')\Bigl\{\bigl[\hat{a},[\hat{\sigma}_+,\hat{\rho}_S]\bigl]+\bigl[\hat{a}^\dagger,[\hat{\sigma}_-,\hat{\rho}_S]\bigl]\Bigl\}\nonumber\\
&-\bar{\Omega}\pi J(\omega_0)\Bigl\{\bigl[\hat{a}^\dagger,\hat{\sigma}_-\hat{\rho}_S\bigl]-\bigl[\hat{a},\hat{\rho}_S\hat{\sigma}_+\bigl]\Bigl\}-\bar{\Omega}\pi J(\omega)\Bigl\{\hat{a}\bigl[\hat{\sigma}_+,\hat{\rho}_S\bigl]\nonumber\\
    &-\bigl[\hat{\sigma}_-,\hat{\rho}_S\bigl]\hat{a}^\dagger\Bigl\}-i\bar{\Omega}\lambda_2'\Bigl\{\bigl[\hat{a}^\dagger,\{\hat{\sigma}_-,\hat{\rho}_S\}\bigl]+\bigl[\hat{a},\{\hat{\sigma}_+,\hat{\rho}_S\}\bigl]\Bigl\}\nonumber\\
    &-i\bar{\Omega}d\Bigl\{\bigl[\hat{a}^\dagger,\hat{\sigma}_-\hat{\rho}_S\bigl]+\bigl[\hat{a},\hat{\rho}_S\hat{\sigma}_+\bigl]\Bigl\}+i\bar{\Omega}\lambda_1'\Bigl\{\bigl[\hat{a},[\hat{\sigma}_+,\hat{\rho}_S]\bigl]\nonumber\\
    &-\bigl[\hat{a}^\dagger,[\hat{\sigma}_-,\hat{\rho}_S]\bigl]\Bigl\}+i\bar{\Omega}d'\Bigl\{\hat{a}\bigl[\hat{\sigma}_+,\hat{\rho}_S\bigl]+\bigl[\hat{\sigma}_-,\hat{\rho}_S\bigl]\hat{a}^\dagger\Bigl\},
\end{align}
\noindent
where,
\begin{align}
    &\alpha_2' = \pi J(\omega_0)n(\omega_0), \hspace{2mm} \bar{\Omega} = \alpha x_0, \hspace{2mm} \alpha_1' = \pi J(\omega)n(\omega),\nonumber\\
    &\lambda_2'=\mathrm{P}\int d\tilde{\omega}\frac{J(\tilde{\omega})n(\tilde{\omega})}{\tilde{\omega}-\omega_0},\hspace{2mm}d=\mathrm{P}\int d\tilde{\omega}\frac{J(\tilde{\omega})}{\tilde{\omega}-\omega_0},\nonumber\\
    &\lambda_1'=\mathrm{P}\int d\tilde{\omega}\frac{J(\tilde{\omega})n(\tilde{\omega})}{\tilde{\omega}-\omega},\hspace{2mm}d'=\mathrm{P}\int d\tilde{\omega}\frac{J(\tilde{\omega})}{\tilde{\omega}-\omega}.
\end{align}

The Born-Markov quantum master equation~(\ref{master_eq}) is of GKSL form~\cite{breuer2002theory,gorini1976completely,lindblad1976generators}, which guarantees completely positive dynamics.

\section{Adiabatic elimination of fast variables}\label{elimination}
\subsection{Coordinate representation}

To show that Eq.~(\ref{master_eq}) can be written in the form of a master equation describing the OQBM~\cite{bauer2013open,bauer2014open}, one needs to show that the master equation~(\ref{master_eq}) can be represented in a purely diagonal representation in position. One achieves this by starting from a generic non-diagonal representation of the form:
\begin{equation}\label{xy_co}
    \hat{\rho}_S (t) = \int_{-\infty}^{+\infty} dx dy \rho(x,y)\otimes\ket{x}\bra{y}.
\end{equation}
\noindent
By substituting Eq.~(\ref{xy_co}) into Eq.~(\ref{master_eq}) and using the relation between the position and momentum eigenstate, $\langle x | p \rangle = e^{ipx/\hbar}/\sqrt{2\pi \hbar}$,  it is straightforward to show that the equation of motion for $\rho(x,y)$ reads
\begin{align}\label{master_eq_xy}
    \frac{\partial}{\partial t} \rho(x,y) = \pazocal{L}_\mathrm{QHO}\rho+\pazocal{L}_\mathrm{2LS}\rho+\pazocal{L}_\mathrm{cross}\rho,
\end{align}
\noindent
where,
\begin{align}\label{me_xy}
&\pazocal{L}_\mathrm{QHO}\rho(x,y)=\Biggl[\frac{i\hbar}{2m}\biggl(\frac{\partial^2}{\partial x^2}-\frac{\partial^2}{\partial y^2}\biggl)-\frac{im\omega^2}{2\hbar}\bigl(x^2-y^2\bigl)\nonumber\\
&-\bar{\alpha}_1(x-y)^2+\bar{\alpha}_2\hbar^2\biggl(\frac{\partial^2}{\partial x^2}+\frac{\partial^2}{\partial y^2}+2\frac{\partial^2}{\partial y\partial x}\biggl)\nonumber\\
&+2\hbar\bar{\alpha}_3\biggl(1+x\frac{\partial}{\partial y}+y\frac{\partial}{\partial x}\biggl)\Biggl]\rho,\\
&\pazocal{L}_\mathrm{2LS}\rho(x,y) = -i\bigl[(\omega_0/2)\hat{\sigma}_z+\Omega\hat{\sigma}_x,\rho\bigl]+i\beta_3\bigl[\hat{\sigma}_z,\rho\bigl]\nonumber\\
&+\beta_1\pazocal{L}[\hat{\sigma}_-,\hat{\sigma}_+]\rho+\beta_2\pazocal{L}[\hat{\sigma}_+,\hat{\sigma}_-]\rho,
\end{align}
and
\begin{align}
&\pazocal{L}_\mathrm{cross}\rho(x,y) = -\alpha_{11}\biggl\{m\omega(x-y)[\hat{\sigma}_x,\rho]+\hbar\biggl(\frac{\partial}{\partial x}+\frac{\partial}{\partial y}\biggl)\nonumber\\
&\times[i\hat{\sigma}_y,\rho]\biggl\}-\alpha_{12}\biggl\{m\omega(x-y)\bigl\{\hat{\sigma}_x\rho-\{\hat{\sigma}_+,\rho\}\bigl\}-\hbar\biggl(\frac{\partial}{\partial x}\nonumber\\
&+\frac{\partial}{\partial y}\biggl)\bigl\{\hat{\sigma}_x\rho-[\hat{\sigma}_+,\rho]\bigl\}\biggl\}-\alpha_{13}\biggl\{m\omega x[\hat{\sigma}_+,\rho]-m\omega y[\hat{\sigma}_-,\rho]\nonumber\\
&+\hbar\frac{\partial}{\partial x}[\hat{\sigma}_+,\rho]-\hbar\frac{\partial}{\partial y}[\hat{\sigma}_-,\rho]\biggl\}-i\alpha_{14}\biggl\{m\omega (x-y)\{\hat{\sigma}_x,\rho\}\nonumber\\
&+\hbar\frac{\partial}{\partial x}\{\hat{i\sigma}_y,\rho\}+\hbar\frac{\partial}{\partial y}\{i\hat{\sigma}_y,\rho\}\biggl\}-i\alpha_{15}\biggl\{m\omega (x-y)\bigl\{\hat{\sigma}_x\rho\nonumber\\
&-[\hat{\sigma}_+,\rho]\bigl\}+\hbar\biggl(\frac{\partial}{\partial x}+\frac{\partial}{\partial y}\biggl)\bigl\{\{\hat{\sigma}_+,\rho\}-\hat{\sigma}_x\rho\bigl\}\biggl\}\nonumber\\
&+i\alpha_{16}\biggl\{m\omega (x-y)[i\hat{\sigma}_y,\rho]+\hbar\biggl(\frac{\partial}{\partial x}+\frac{\partial}{\partial y}\biggl)[\hat{\sigma}_x,\rho]\biggl\}\nonumber\\
&+i\alpha_{17}\biggl\{m\omega x[\hat{\sigma}_+,\rho]+m\omega y[\hat{\sigma}_-,\rho]+\hbar\frac{\partial}{\partial x}[\hat{\sigma}_+,\rho]\nonumber\\
&+\hbar\frac{\partial}{\partial y}[\hat{\sigma}_-,\rho]\biggl\}.\label{crxy}
\end{align}
\noindent
The parameters appearing in Eq.~(\ref{crxy}) are defined as
\begin{align}
&\alpha_{11}=\alpha_0(\alpha_2'+\alpha_1'), \hspace{4mm} \alpha_{12}=\bar{\Omega}J(\omega_0)\pi\alpha_0,\nonumber\\
&\alpha_{13}=\bar{\Omega}J(\omega)\pi\alpha_0,
\hspace{4mm} \alpha_{14}=\bar{\Omega}\lambda_2'\alpha_0,\hspace{4mm}\alpha_{15}=\bar{\Omega}d\alpha_0,\nonumber\\
&\alpha_{16}=\bar{\Omega}\lambda_1'\alpha_0,\hspace{4mm} \alpha_{17}=\bar{\Omega}d'\alpha_0,\hspace{4mm}\alpha_0 = 1/\sqrt{2\hbar m\omega}.
\end{align}
\noindent
Subsequently, we make a coordinate transformation to the canonical form, which makes the adiabatic elimination of fast variables straightforward via
\begin{equation}     
u = \frac{x+y}{2}, \hspace{4mm} v = x - y. 
\end{equation} 
\noindent 
In the rotated coordinate, Eq.~(\ref{master_eq_xy}) becomes
\begin{align}\label{uv_eq}
    \frac{\partial}{\partial t} \rho(u,v) = \pazocal{L}_\mathrm{QHO}\rho+\pazocal{L}_\mathrm{2LS}\rho+\pazocal{L}_\mathrm{cross}\rho,
\end{align}
\noindent
where,
\begin{align}\label{uv_eqf}
&\pazocal{L}_\mathrm{QHO}\rho(u,v)=\biggl[\frac{i\hbar}{m}\frac{\partial^2}{\partial v \partial u}-\frac{im\omega^2}{\hbar}uv - \bar{\alpha}_1 v^2+2\hbar\bar{\alpha}_3\nonumber\\
&+\bar{\alpha}_2\hbar^2\frac{\partial^2}{\partial u^2}+2\hbar\bar{\alpha}_3 u \frac{\partial}{\partial u}-2\hbar\bar{\alpha}_3 v \frac{\partial}{\partial v}\biggl]\rho,\\
&\pazocal{L}_\mathrm{2LS}\rho(u,v)= -i\bigl[(\omega_0/2)\hat{\sigma}_z+\Omega\hat{\sigma}_x,\rho\bigl]+i\beta_3\bigl[\hat{\sigma}_z,\hat{\rho}\bigl]\nonumber\\
    &+\beta_1\pazocal{L}[\hat{\sigma}_-,\hat{\sigma}_+]\rho+\beta_2\pazocal{L}[\hat{\sigma}_+,\hat{\sigma}_-]\rho,\\
    &\pazocal{L}_\mathrm{cross}\rho(u,v)= \biggl(\frac{\partial}{\partial u} \hat{m}_1+\frac{\partial}{\partial v} \hat{m}_2+ u \hat{m}_3
   +v \hat{m}_4\biggl)\rho. \label{crossp}
\end{align}
\noindent
The super-operators acting on the internal degree of freedom, namely $\hat{m}_1, \hat{m}_2, \hat{m}_3$, and $\hat{m}_4$, are respectively given by
\begin{align}
    &\hat{m}_1 = -\tilde{a}_1\bigl[i\hat{\sigma}_y,\cdot\bigl]+\tilde{a}_2\Bigl\{\hat{\sigma}_x\cdot-\bigl[\hat{\sigma}_+,\cdot\bigl]\Bigl\}-\tilde{a}_3\bigl[i\hat{\sigma}_y,\cdot\bigl]\nonumber\\
    &-i\tilde{a}_4\big\{i\hat{\sigma}_y,\cdot\bigl\}-i\tilde{a}_5\Bigl\{\bigl\{\hat{\sigma}_+,\cdot\bigl\}-\hat{\sigma}_x\cdot\Bigl\}+i\tilde{a}_6\bigl[\hat{\sigma}_x,\cdot\bigl],\nonumber\\
    &\hat{m}_2= -\tilde{a}_1\bigl[i\hat{\sigma}_y,\cdot\bigl]+\tilde{a}_2\Bigl\{\hat{\sigma}_x\cdot-\bigl[\hat{\sigma}_+,\cdot\bigl]\Bigl\}-\tilde{a}_3\bigl[i\hat{\sigma}_y,\cdot\bigl]\nonumber\\
    &\hat{m}_3 = i\tilde{a}_7\bigl[\hat{\sigma}_x,\cdot\bigl]-i\tilde{a}_8\bigl[\hat{\sigma}_y,\cdot\bigl],\nonumber\\
     &\hat{m}_4 = -\frac{1}{2}(2\tilde{a}_1+\tilde{a}_3)\bigl[\hat{\sigma}_x,\cdot\bigl]-\tilde{a}_2\Bigl\{\hat{\sigma}_x\cdot-\bigl\{\hat{\sigma}_+,\cdot\bigl\}\Bigl\}\nonumber\\
     &-i\tilde{a}_4\bigl\{\hat{\sigma}_x,\cdot\bigl\}-i\tilde{a}_5\Bigl\{\hat{\sigma}_x\cdot-\bigl[\hat{\sigma}_+,\cdot\bigl]\Bigl\}+\frac{i}{2}(2\tilde{a}_6+\tilde{a}_7)\bigl[i\hat{\sigma}_y,\cdot\bigl],
\end{align}
\noindent
where,
\begin{align}
    \tilde{a}_1 &= \hbar\pi\alpha_0(J(\omega_0)n(\omega_0)+J(\omega)n(\omega)),\hspace{2mm}\tilde{a}_2=\frac{\hbar\pi\alpha}{2m\omega}J(\omega_0),\nonumber\\
\tilde{a}_3&=\frac{\hbar\pi\alpha}{4m\omega}J(\omega), \hspace{2mm} \tilde{a}_4=\frac{\hbar\alpha}{2m\omega}\mathrm{P}\int d\tilde{\omega} \frac{J(\tilde{\omega})n(\tilde{\omega})}{\tilde{\omega}-\omega_0},\nonumber\\
\tilde{a}_5&=\frac{\hbar\alpha}{2m\omega}\mathrm{P}\int d\tilde{\omega} \frac{J(\tilde{\omega})}{\tilde{\omega}-\omega_0},\hspace{2mm}\tilde{a}_7=\frac{\alpha}{2}\mathrm{P}\int d\tilde{\omega} \frac{J(\tilde{\omega})}{\tilde{\omega}-\omega},\nonumber\\
\tilde{a}_6&=\frac{\hbar\alpha}{2m\omega}\mathrm{P}\int d\tilde{\omega} \frac{J(\tilde{\omega})(n(\tilde{\omega})+1/2)}{\tilde{\omega}-\omega},\hspace{2mm}\tilde{a}_8=\frac{\alpha J(\omega)}{2}\pi.
\end{align}
\noindent
Above, $\mathrm{P}$ denotes the Cauchy principal value. In the following subsection, we perform adiabatic elimination of fast variables and derive a completely positive master equation for the diagonal elements that describes OQBM.

\subsection{Elimination of fast variables}

Here, we present a systematic method for eliminating fast-evolving variables from a distribution, and derive an equation of motion that describes the dynamics of the slow variables~\cite{van1985elimination,gardiner1985handbook,kramers1940brownian}. This method is well established and was central to deriving the Smoluchowski equation~\cite{smoluchowski1916brownsche}. The system~(\ref{uv_eq}) under consideration involves two variables, $u$ and $v$, with distribution $\rho(u,v)$, where $u$ denotes the slow variable and $v$ the fast variable, with the latter being the one we eliminate. As a first step, we rewrite Eq.~(\ref{uv_eq}) as
\begin{align}\label{weqfin3}
   &\frac{\partial}{\partial t}\rho(u,v) =\Bigl(\bar{\alpha}_1 \hat{L}_1 +\hat{L}_2\Bigl)\rho+\biggl(\bar{\alpha}_2\hbar^2\frac{\partial^2}{\partial u^2}+2\bar{\alpha}_3\hbar\nonumber\\
   &+2\bar{\alpha}_3\hbar u\frac{\partial}{\partial u}\biggl)\rho+\biggl(\frac{\partial}{\partial u} \hat{m}_1+\frac{\partial}{\partial v} \hat{m}_2+ u \hat{m}_3
   +v \hat{m}_4\biggl)\rho \nonumber\\
   &+\pazocal{L}_{\mathrm{2LS}} \rho,
\end{align}
\noindent
where $\hat{L}_1$ and $\hat{L}_2$, are 
  \begin{align}
     \hat{L}_1 & = -v^2-\gamma v \frac{\partial}{\partial v}, \label{eq_1}\\
    \hat{L}_2 & = \frac{i\hbar}{m}\frac{\partial^2}{\partial v \partial u}-\frac{i}{\hbar}m\omega^2uv. \label{eq_2d}
\end{align}
\noindent
Above, $\gamma = 2\hbar \bar{\alpha}_3 /\bar{\alpha}_1 $. We assume that $(m\omega)^2\sim k_BT$ and that $\bar{\alpha}_1$ is significantly larger than all the system parameters, which is the time scale for the fast variable. Our goal is to derive a position distribution function $\bar{\rho}(u) = \int_{-\infty}^{+\infty} dv \rho(u,v)$ valid in the limit $\bar{\alpha}_1 \gg 1$. 
After performing the adiabatic elimination of fast variables~\cite{smoluchowski1916brownsche,van1985elimination,gardiner1985handbook,kramers1940brownian} (see the Appendix~\ref{appendixa} for details), we find that $v=0$, which implies that $u=x$, and we derive a completely positive master equation for the diagonal elements, which reads 
\begin{align}\label{oqbm_me}
    \frac{\partial}{\partial t} \bar{\rho}(x,t) &\approx -\lambda_1x\frac{\partial}{\partial x }\bar{\rho}-\lambda_2 x^2 \bar{\rho}+\hat{m}_1 \frac{\partial}{\partial x }\bar{\rho}+x\hat{m}_3\bar{\rho}\nonumber\\
    &\quad+\lambda_3\frac{\partial^2}{\partial x^2 }\bar{\rho}+\lambda_4 \frac{\partial}{\partial x }(x\bar{\rho} )+ \pazocal{L}_\mathrm{2LS}\bar{\rho},
    \end{align}
\noindent
where,
\begin{align}
    \lambda_1=\frac{\omega^2}{\bar{\alpha}_1\gamma}, \hspace{2mm} \lambda_2=\frac{m^2\omega^4}{\bar{\alpha}_1\hbar^2},\hspace{2mm} \lambda_3=\bar{\alpha}_2\hbar^2, \hspace{2mm} \lambda_4=2\bar{\alpha}_3\hbar.
\end{align} 
The above master equation~(\ref{oqbm_me}) defines OQBM. It has the same structure as the original master equation derived by Bauer \textit{et al.} (see Eq. (2) in~\cite{bauer2013open}, and Eq. (28) in~\cite{bauer2014open}) and from the microscopic derivation of OQBM by~\cite{sinayskiy2015microscopicbrown,sinayskiy2017steady} (see Eq. (13) on both papers) and also Eq.~(81) in our recent paper~\cite{zungu2025adiabatic}.

The propagation of the Brownian particle is governed by the diffusive term ($-\lambda_1x\frac{\partial}{\partial x }\bar{\rho}-\lambda_2 x^2 \bar{\rho}+\lambda_3\frac{\partial^2}{\partial x^2 }\bar{\rho}+\lambda_4\frac{\partial}{\partial x}(x\bar{\rho})$) in the master equation~(\ref{oqbm_me}). 
The Lindblad term ($\pazocal{L}_\mathrm{2LS}\bar{\rho}$) denotes the dissipative dynamics of the internal degree of freedom of the  Brownian particle.
The remaining term ($\hat{m}_1 \frac{\partial}{\partial x }\bar{\rho}+x\hat{m}_3\bar{\rho}$) couples the external and internal degrees of freedom via the bosonic bath.

\section{Numerical illustrations of OQBM dynamics}\label{numerical_examples}

The reduced density matrix~$\bar{\rho}(x,t)$ of the open quantum Brownian particle~(\ref{oqbm_me}) can be written in the form:
\begin{equation}\label{matrix}
  \bar{\rho}(x,t) = \begin{pmatrix}
\rho_{1,1}(x,t) & \rho_{1,2}(x,t) \\
\rho_{2,1}(x,t) & \rho_{2,2}(x,t)
\end{pmatrix}.	
\end{equation}
\noindent
The diagonal elements $\rho_{1,1}(x,t)$ and $\rho_{2,2}(x,t)$ represent the populations in the two quantum states. The off-diagonal elements $\rho_{1,2}(x,t)=(\rho_{2,1}(x,t))^*$ represent the coherences, which encode the quantum superposition and phase relationships (here ($^*$) denotes complex conjugate).

\begin{figure}[htbp]
\centering 
{%
\includegraphics[width=8.6cm]{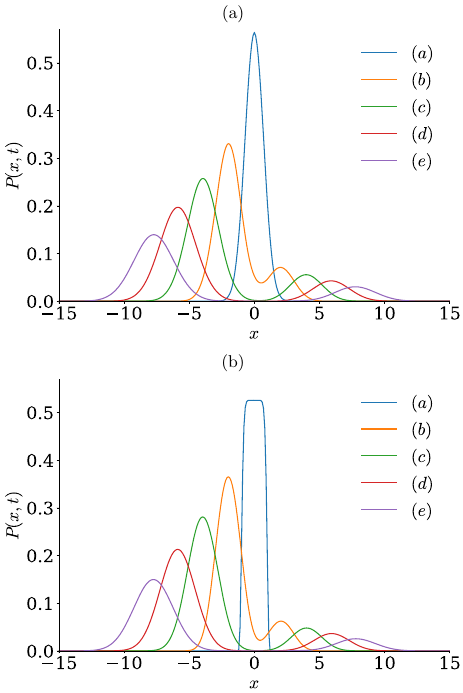}%
}
\caption{The probability distribution $P(x,t)$ of the open quantum Brownian particle at position 
$x$ at various times.  Curves ($a$) to ($e$) correspond to times 0, 50, 100, 150, and 200, respectively.
The initial distribution is given by Eq.~(\ref{fn_1}) with $j=2$, $\theta=\pi/5$, $\phi=\pi/4$ corresponding to subplot (a); and $j=10$, $\theta=\pi/4$, $\phi=\pi/4$ corresponding to subplot (b). Other parameters are $\Omega=0.5$, $\beta=10^{-4}$, $\Delta_1=10^{-5}$, $\Delta_2=0.01$, $\Delta_3=10^{-4}$, $\Delta_4=8\times10^{-3}$, $\lambda_2=10^{-4}$, $\lambda_3=5\times10^{-3}$,  $\lambda_4=10^{-4}$, $\delta_1=10^{-4}$, $\delta_2=0.04$, $\delta_3=0.01$, $\tilde{a}_2=0.04$, $\tilde{a}_7=10^{-3}$, and $\tilde{a}_8=10^{-4}$.}
\label{example1}
\end{figure}

Using the above form of the density matrix~(\ref{matrix}), the master equation~(\ref{oqbm_me}) can be rewritten as system of partial differential equations:
\begin{align}\label{system_pde}
    \frac{\partial}{\partial t} &\rho_+ = \biggl\{\lambda_3 \frac{\partial^2}{\partial x^2} +\Delta_1 x  \frac{\partial}{\partial x} +\lambda_4 -\lambda_2 x^2 \biggl\} \rho_+ \nonumber\\
   &- 2 \delta_1 \frac{\partial}{\partial x} C_I+2 \tilde{a}_2 \frac{\partial}{\partial x} C_R,\nonumber\\
   \frac{\partial}{\partial t} &\rho_- = \biggl\{\lambda_3 \frac{\partial^2}{\partial x^2} +\Delta_1 x  \frac{\partial}{\partial x} +\Delta_2 -\lambda_2 x^2 \biggl\} \rho_--\bar{\beta}\rho_+\nonumber\\
   &+\biggl\{4\tilde{a}_7x-4\Omega+4\delta_2\frac{\partial}{\partial x}\biggl\}C_I-\biggl\{4\tilde{a}_8x+4\delta_3\frac{\partial}{\partial x}\biggl\}C_R,\nonumber\\
   \frac{\partial}{\partial t} &C_R = \biggl\{\lambda_3 \frac{\partial^2}{\partial x^2} +\Delta_1 x  \frac{\partial}{\partial x} +\Delta_3 -\lambda_2 x^2 \biggl\} C_R\nonumber\\
   &+\biggl\{\tilde{a}_8x+\delta_3\frac{\partial}{\partial x}\biggl\}\rho_-+\frac{1}{2}\tilde{a}_2\frac{\partial}{\partial x}\rho_+,\nonumber\\
   \frac{\partial}{\partial t} &C_I = \biggl\{\lambda_3 \frac{\partial^2}{\partial x^2} +\Delta_1 x  \frac{\partial}{\partial x} +\Delta_4 -\lambda_2 x^2 \biggl\} C_I\nonumber\\
   &+\biggl\{\Omega-\tilde{a}_7x-\delta_2\frac{\partial}{\partial x}\biggl\}\rho_--\frac{1}{2}\delta_1\frac{\partial}{\partial x}\rho_+,
\end{align}
\noindent
where $\rho_\pm = \rho_{1,1}(x,t)\pm \rho_{2,2}(x,t)$, $C_R = \text{Re}(\rho_{1,2}(x,t))$, $C_I = \text{Im}(\rho_{1,2}(x,t))$ and the parameters are given by
\begin{align}
    &\Delta_1= \lambda_4 -\lambda_1, \hspace{2mm} \Delta_2= \lambda_4 - \beta_2-\beta_1, \hspace{2mm} \bar{\beta}=\beta_1-\beta_2, \nonumber\\
    &\Delta_3= \lambda_4 - \frac{1}{2}(\beta_2+\beta_1),\hspace{2mm}\delta_1=2\tilde{a}_4+\tilde{a}_5,\hspace{2mm}\delta_2=\frac{1}{2}\tilde{a}_5+\tilde{a}_6,\nonumber\\
    &\Delta_4= 2\beta_3+\Delta_3-\omega_0,\hspace{2mm}\delta_3=\tilde{a}_1+\tilde{a}_3+\frac{1}{2}\tilde{a}_2.
\end{align}
\noindent
To illustrate the OQBM dynamics, we numerically integrate Eq.~(\ref{system_pde}) for various system-bath parameters and boundary conditions. We explore Gaussian ($j=2$) and non-Gaussian (e.g., $j=10$) initial distributions for the quantum Brownian particle, and we assume that the internal degree of freedom is initially a pure state, which reads:
\begin{align}\label{fn_1}
  \bar{\rho}_j(x,0) = \frac{1}{2I_j}e^{-x^j}\otimes\begin{pmatrix}
2\cos^2\theta & \sin 2\theta e^{-i\phi} \\
\sin 2\theta e^{i\phi} & 2\sin^2\theta
\end{pmatrix},	
\end{align}
\noindent
where $I_j=\int_{-\infty}^{+\infty} dx e^{-x^j}$,  $\theta\in [0,\pi)$, $\phi\in [0,2\pi)$, and $j>0$. 
Figure~\ref{example1} depicts the position probability distribution $P(x, t) = \tr (\rho_+(x,t))$ of finding the open quantum Brownian particle at certain positions for different moments of time. In Fig.~\ref{example1}(a), we chose a Gaussian distribution ($j=2$) as the initial position distribution of the OQBM walker, and we chose a mixed state with partial coherences for the internal degree of freedom.
In this example, coherence from the internal degree of freedom is transferred to the spatial dynamics. This causes the initial single Gaussian distribution to split into two Gaussians that propagate at different rates. For $t\ge50$, the distribution becomes skewed to the left, depending on the system-bath parameters.

The asymmetric spreading results from the choice of the initial state of the internal degree of freedom ($\theta$ and $\phi$).
The number of Gaussian profiles at time $t\ge100$ is not limited to two Gaussian profiles, demonstrating that OQBM dynamics are richer than OQWs~\cite{attal2012open, ATTAL20121545} dynamics. The probability distribution of this OQBM model propagates significantly in both the left and right directions for various parameters, exhibiting a different spread compared to the OQBM described in~\cite{sinayskiy2015microscopicbrown,sinayskiy2017steady,zungu2025adiabatic}.

Figure~\ref{example1}(b) shows the probability distribution evolving to a Gaussian form even from a non-Gaussian initial state (for $j=10$).  In both examples, Fig.~\ref{example1}(a)-(b), the probability distribution at times $t\ge50$, clearly approaches a Gaussian distribution and visibly splits into two Gaussian profiles propagating to the left and right directions.

\begin{figure}[htbp]
\centering 
{%
\includegraphics[width=8.6cm]{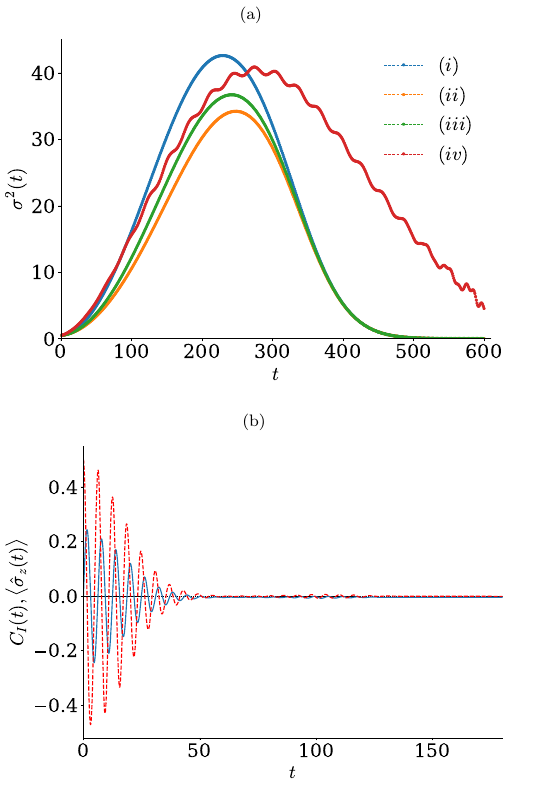}%
}
\caption{Illustration of OQBM dynamics. 
Subplot (b) shows the variance $\sigma^2(t)$ as a function of time $t$ for different OQBM probability distributions. The initial distribution is given by Eq.~(\ref{fn_1}) with $j=2$. Curve ($i$) corresponds to the parameters, $\theta=\pi/2$, $\phi=\pi/4$, $\Omega=0.5$, $\beta=10^{-3}$, $\Delta_1=10^{-5}$, $\Delta_2=10^{-3}$, $\Delta_3=10^{-4}$, $\Delta_4=8\times10^{-3}$, $\lambda_2=10^{-4}$, $\lambda_3=5\times10^{-3}$,  $\lambda_4=10^{-4}$, $\delta_1=10^{-4}$, $\delta_2=0.04$, $\delta_3=0.01$, $\tilde{a}_2=0.04$, $\tilde{a}_7=10^{-3}$, and $\tilde{a}_8=10^{-4}$; Curve ($ii$) corresponds to Fig.~\ref{example1}(b); Curve ($iii$) correspond to $\theta=\pi/6$, $\phi=\pi/4$, $\Omega=0.5$, $\beta=10^{-4}$, $\Delta_1=10^{-5}$, $\Delta_2=10^{-3}$, $\Delta_3=10^{-4}$, $\Delta_4=8\times10^{-3}$, $\lambda_2=10^{-4}$, $\lambda_3=5\times10^{-3}$,  $\lambda_4=10^{-4}$, $\delta_1=10^{-4}$, $\delta_2=0.04$, $\delta_3=0.01$, $\tilde{a}_2=-0.04$, $\tilde{a}_7=10^{-3}$, and $\tilde{a}_8=10^{-4}$; Curve ($iv$) correspond to $\theta=\pi$, $\phi=0$, $\Omega=0.1$, $\beta=0.01$, $\Delta_1=10^{-5}$, $\Delta_2=10^{-3}$, $\Delta_3=2\times10^{-4}$, $\Delta_4=0.01$, $\lambda_2=10^{-4}$, $\lambda_3=0.01$,  $\lambda_4=10^{-3}$, $\delta_1=0.01$, $\delta_2=0.02$, $\delta_3=0.03$, $\tilde{a}_2=0.04$, $\tilde{a}_7=10^{-3}$, and $\tilde{a}_8=0.02$. Subplot (b) shows the time evolution of the imaginary part of the off-diagonal element $\bigl(C_I (t) = {\tr}_x\bigl[C_I(x,t)\bigl]\bigl)$ (solid curve) and the expectation value of $\langle \hat{\sigma}_z (t) \rangle$ (dashed curve) of the open quantum Brownian particle. The parameters are $\theta=\pi/6$, $\phi=0$, $\Omega=0.5$, $\beta=0.01$, $\Delta_1=10^{-4}$, $\Delta_2=10^{-3}$, $\Delta_3=10^{-4}$, $\Delta_4=8\times10^{-3}$, $\lambda_2=10^{-4}$, $\lambda_3=4\times10^{-3}$,  $\lambda_4=10^{-4}$, $\delta_1=0.01$, $\delta_2=0.06$, $\delta_3=0.01$, $\tilde{a}_2=10^{-3}$, $\tilde{a}_7=0.04$, and $\tilde{a}_8=10^{-4}$.}
\label{example2}
\end{figure}

Figure~\ref{example2}(a) shows the time-dependent variance $\sigma^2(t)$ in the position of the OQBM walker. Curves $(i)$ through $(iv)$ illustrate that $\sigma^2(t)$ grows exponentially due to the coherent driving of the internal degree of freedom of the OQBM walker. However, $\sigma^2(t)$ decreases rapidly after time $t\approx 200$ due to the decoherent interaction with the thermal bath. There appears to be a regime switch at time $t\approx 200$ as the OQBM walker relaxes towards equilibrium. Specifically, in curve ($i$) of Fig.~\ref{example2}(a), we initialized the internal degree of freedom of the OQBM walker with zero coherences. In curves ($ii$)-($iii$), we initialized the internal degree of freedom of the OQBM walker with non-zero coherences. For these curves ($i$) to ($iii$), the drive is set to $\Omega=0.5$, and due to the stronger drive, we see a smoother variance evolution, approaching a Gaussian-like profile. However, for curve ($iv$), we initialized the internal degree of freedom of the OQBM walker with zero coherences, and $\Omega =0.1$. Due to slow driving in the curve ($iv$), the coherences are driven more slowly and persist longer against dissipation. Figure~\ref{example2}(a) thus demonstrates how the variance evolution depends on the initial internal state and the driving strength $\Omega$. These quantities determine whether the Gaussian-like variance evolves smoothly or develops coherent oscillations (``wiggling'') before relaxing.

Figure~\ref{example2}(b) (solid curve) shows the time evolution of the imaginary part of the off-diagonal element of the OQBM density matrix $\bigl(C_I (t) = {\tr}_x\bigl[C_I(x,t)\bigl]\bigl)$. Figure~\ref{example2}(b) (dashed curve) shows the time evolution of the inverse population in the quantum internal degree of freedom of the open Brownian particle $\bigl(\langle \hat{\sigma}_z (t) \rangle=\tr(\hat{\sigma}_z\bar{\rho}(x,t))\bigl)$. 
As shown in Fig.~\ref{example2}(b), both $C_I (t)$ and $\langle \hat{\sigma}_z (t)\rangle$ exhibit initial coherence before decaying to zero due to interaction with the bath. However, at time $t\approx100$, both $C_I (t)$ and $\langle \hat{\sigma}_z (t) \rangle$ exhibit small-amplitude revivals in the evolution, which reflect partial re-coherence effects in the system before the thermal bath fully suppresses them.

\section{Moments and cumulant generating functions}\label{moments_cumulants}
\subsection{Equations for the moments of the position distribution}\label{moment_c}

To further investigate the dynamical behavior of the OQBM walker, we now derive the $n$-th moments of its position distribution. To simplify the derivation, we first nondimensionalize the system of partial differential equations~(\ref{system_pde}) by introducing the dimensionless coordinate $\xi=x/\tilde{x}_0$ and rescaling the density functions accordingly, i.e., $\tilde{\rho}_\pm = \tilde{x}_0 \rho_\pm$, $\tilde{C}_{R,I} = \tilde{x}_0 C_{R,I}$. We denote the $n$-th moment of the scaled distribution by $\tilde{\rho}(\xi,t)$, i.e.,
\begin{equation}\label{momentss}
    \bigl\langle \xi^n \tilde{\rho}(\xi,t)\bigl\rangle = \int_{-\infty}^{+\infty}  d\xi \xi^n \tilde{\rho}(\xi, t),
\end{equation}
\noindent
where $\tilde{\rho}(\xi,t)\equiv\bigl\{\tilde{\rho}_\pm,\tilde{C}_{R,I}\bigl\}$. By directly inserting Eq.~(\ref{momentss}) into Eq.~(\ref{system_pde}), we obtain
\begin{align}\label{pde_nth}
    &\frac{d}{dt} \bigl\langle \xi^n \tilde{\rho}_+\bigl\rangle = \bar{\lambda}_3n(n-1) \bigl\langle \xi^{n-2}\tilde{\rho}_+ \bigl\rangle+[\lambda_4-\Delta_1(n+1)]\nonumber\\
&\times\bigl\langle\xi^n\tilde{\rho}_+\bigl\rangle -\bar{\lambda}_2\bigl\langle\xi^{n+2}\tilde{\rho}_+\bigl\rangle+2\bar{\delta}_1n\bigl\langle\xi^{n-1}\tilde{C}_I\bigl\rangle-2\bar{a}_2n\bigl\langle\xi^{n-1}\tilde{C}_R\bigl\rangle,\nonumber\\
     &\frac{d}{dt} \bigl\langle \xi^n \tilde{\rho}_-\bigl\rangle = \bar{\lambda}_3n(n-1) \bigl\langle \xi^{n-2}\tilde{\rho}_- \bigl\rangle+[\Delta_2-\Delta_1(n+1)]\nonumber\\
&\times\bigl\langle\xi^n\tilde{\rho}_-\bigl\rangle -\bar{\lambda}_2\bigl\langle\xi^{n+2}\tilde{\rho}_-\bigl\rangle-\bar{\beta}\bigl\langle\xi^n\tilde{\rho}_+\bigl\rangle+4\bar{a}_7\bigl\langle\xi^{n+1}\tilde{C}_I\bigl\rangle\nonumber\\
&-4\Omega\bigl\langle\xi^n\tilde{C}_I\bigl\rangle+4\bar{\delta}_2n\bigl\langle\xi^{n-1}\tilde{C}_I\bigl\rangle-4\bar{a}_8\bigl\langle\xi^{n+1}\tilde{C}_R\bigl\rangle\nonumber\\
&+4\bar{\delta}_3n\bigl\langle\xi^{n-1}\tilde{C}_R\bigl\rangle,\nonumber\\
&\frac{d}{dt} \bigl\langle \xi^n \tilde{C}_R\bigl\rangle = \bar{\lambda}_3n(n-1) \bigl\langle \xi^{n-2}\tilde{C}_R \bigl\rangle+[\Delta_3-\Delta_1(n+1)]\nonumber\\
&\times\bigl\langle\xi^n\tilde{C}_R\bigl\rangle-\bar{\lambda}_2\bigl\langle\xi^{n+2}\tilde{C}_R\bigl\rangle+\bar{a}_8\bigl\langle\xi^n\tilde{\rho}_-\bigl\rangle-\bar{\delta}_3n\bigl\langle\xi^{n-1}\tilde{\rho}_-\bigl\rangle\nonumber\\
&-\frac{\bar{a}_2}{2}n\bigl\langle\xi^{n-1}\tilde{\rho}_+\bigl\rangle,\nonumber\\
&\frac{d}{dt} \bigl\langle \xi^n \tilde{C}_I\bigl\rangle = \bar{\lambda}_3n(n-1) \bigl\langle \xi^{n-2}\tilde{C}_I \bigl\rangle+[\Delta_4-\Delta_1(n+1)]\nonumber\\
&\times\bigl\langle\xi^n\tilde{C}_I\bigl\rangle-\bar{\lambda}_2\bigl\langle\xi^{n+2}\tilde{C}_I\bigl\rangle+\Omega\bigl\langle\xi^n\tilde{\rho}_-\bigl\rangle-\bar{a}_7\bigl\langle\xi^{n+1}\tilde{\rho}_-\bigl\rangle\nonumber\\
&+\bar{\delta}_2n\bigl\langle\xi^{n-1}\tilde{\rho}_-\bigl\rangle+\frac{\bar{\delta}_1}{2}n\bigl\langle\xi^{n-1}\tilde{\rho}_+\bigl\rangle.
\end{align}
\noindent
The system~(\ref{pde_nth}) forms a linear infinite-dimensional set of coupled differential equations governing the $n$-th moments, and the parameters are 
\begin{align}
    \bar{\lambda}_2 &= \lambda_2\tilde{x}_0^2,\hspace{2mm} \bar{\lambda}_3 = \frac{\lambda_3}{\tilde{x}_0^2}, \hspace{2mm} \bar{\delta}_1 = \frac{\delta_1}{\tilde{x}_0}, \hspace{2mm}\bar{\delta}_2 = \frac{\delta_2}{\tilde{x}_0},\nonumber\\
   \bar{\delta}_3 &= \frac{\delta_3}{\tilde{x}_0},\hspace{2mm} \bar{a}_2 = \frac{\tilde{a}_2}{\tilde{x}_0},
\hspace{2mm}\bar{a}_7=\tilde{a}_7\tilde{x}_0,\hspace{2mm}\bar{a}_8=\tilde{a}_8\tilde{x}_0.
\end{align}
\noindent
To facilitate numerical integration, we recast the system~(\ref{pde_nth}) in the following form:
\begin{align}\label{systemsf}
    \frac{d}{dt}\vec{R}_n &= \hat{M}_n \vec{R}_n + \hat{A}_n \vec{R}_{n-1}+\hat{B}_n\vec{R}_{n-2}+\hat{C}\vec{R}_{n+1}\nonumber\\
    &+\hat{D}\vec{R}_{n+2},
\end{align}
\noindent
where $\vec{R}_n$, $ \vec{R}_{n-1}$, $ \vec{R}_{n-2}$, $\vec{R}_{n+1}$, and $ \vec{R}_{n+2}$ represents column vectors containing the moments components at orders $n$, $n-1$, $n-2$, $n+1$, and $n+2$, respectively, defined as follows
\begin{equation}
    \vec{R}_{n + i} = \begin{pmatrix}
        \langle \xi^{n+i} \tilde{\rho}_+\rangle  \\
        \langle \xi^{n+i} \tilde{\rho}_-\rangle  \\
        \langle \xi^{n+i} \tilde{C}_R\rangle \\
        \langle \xi^{n+i} \tilde{C}_I\rangle 
    \end{pmatrix}, \hspace{4mm} i=0,\pm 1,\pm2.
\end{equation}
\noindent
The four-by-four matrices denoted by $\hat{M}_n, \hat{A}_n$, and $\hat{B}_n$, depend on the index $n$, while $\hat{D}$ and $\hat{C}$ are four-by-four constant matrices, defined below as
\begin{align}
    \hat{M}_n &=  \begin{pmatrix}
       \Delta_4-\Delta_1\tilde{n}&0&0&0 \\
        -\bar{\beta}&\Delta_2-\Delta_1\tilde{n}&0&-4\Omega \\
        0&\bar{a}_8&\Delta_3-\Delta_1\tilde{n}&0 \\
         0&\Omega&0&\Delta_4-\Delta_1\tilde{n}\\
    \end{pmatrix},\nonumber\\
    \hat{A}_n &= n\begin{pmatrix}
       0&0&-2\bar{a}_2&2\bar{\delta}_1\\
    0&0&4\bar{\delta}_3& 4\bar{\delta}_2\\
        -\frac{1}{2}\bar{a}_2&-\bar{\delta}_3&0&0 \\
        \frac{1}{2}\bar{\delta}_1&\bar{\delta}_2&0&0\\
    \end{pmatrix},\hspace{2mm} \hat{D} =-\bar{\lambda}_2\hat{I}_{4\times4},\nonumber\\
    \hat{C} & =  \begin{pmatrix}
       0&0&0&0 \\
        0&0&-4\tilde{a}_8&4\tilde{a}_7 \\
        0&0&0&0 \\
         0&-\tilde{a}_7&0&0\\
    \end{pmatrix},\hspace{1mm} \hat{B}_n =\bar{\lambda}_3n(n-1)\hat{I}_{4\times4},
\end{align}
\noindent
where $\tilde{n}=n+1$.
To solve the equations for the $n$-th moments~(\ref{systemsf}), we consider a Gaussian initial position distribution for the open Brownian particle and assume that its internal degree of freedom is initially a pure state described by
\begin{equation}\label{iniit}
  \tilde{\rho}(\xi,0) = \frac{1}{2\sqrt{\pi}}e^{-\xi^2}\otimes\begin{pmatrix}
2\cos^2\theta & \sin 2\theta e^{-i\phi} \\
\sin 2\theta e^{i\phi} & 2\sin^2\theta
\end{pmatrix},
\end{equation}
\noindent
where $\theta\in [0,\pi)$, and $\phi\in [0,2\pi)$. Using Eq.~(\ref{momentss}), one can show that at time $t=0$,  the arbitrary initial conditions of the $n$-th moments are:
\begin{align}\label{initial_moments}
   \bigl\langle \xi^n \tilde{\rho}_\pm\bigl\rangle &=   \frac{1}{\sqrt{\pi}}\Gamma\biggl(\frac{1+n}{2}\biggl)\tilde{\rho}_\pm,\nonumber\\
      \bigl\langle \xi^n \tilde{C}_{R,I}\bigl\rangle &=   \frac{1}{\sqrt{\pi}}\Gamma\biggl(\frac{1+n}{2}\biggl)\tilde{C}_{R,I}.
\end{align}
\noindent
Under these initial conditions~(\ref{initial_moments}), solutions corresponding to odd integers $n$ are zero, while even integers $n$ yield nonzero values.

\begin{figure}[htbp]
\centering 
{%
\includegraphics[width=8.6cm]{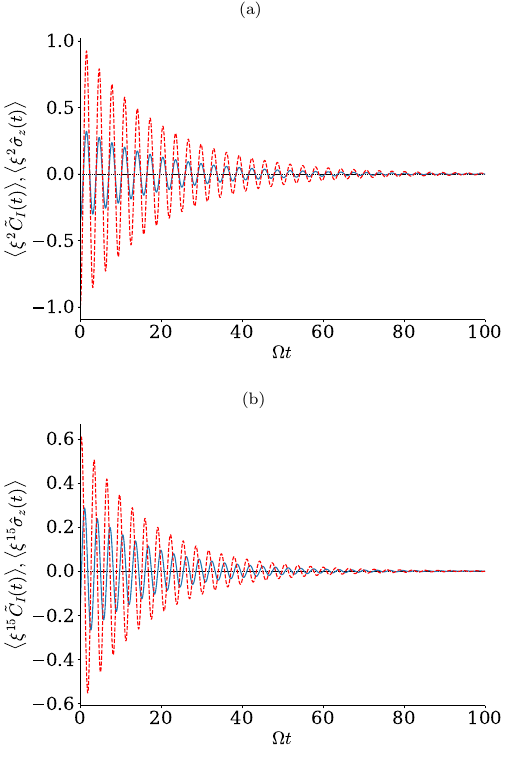}%
}
\caption{The time evolution of the $n$-th moments as a function of dimensionless time: the $n$-th moment with respect to the imaginary part of the off-diagonal element of the OQBM density matrix, $\langle \xi^n \tilde{C}_I(t) \rangle$ (solid curve), and the $n$-th moment of the internal degree of freedom of the open quantum Brownian particle, $\langle \xi^n \hat{\sigma}_z(t) \rangle$ (dashed curve). The initial distribution is given by Eq.~(\ref{iniit}). Subplot (a) corresponds to the second-order moment $n=2$, and $\theta=\phi=\pi/2$, while subplot (b) corresponds to the fifteenth-order moment  $n=15$, and $\theta=\phi=\pi/6$. Other parameters are $\Omega=0.5$, $\bar{\beta}=\Delta_3=\Delta_4=\bar{\lambda}_2=\bar{\lambda}_3=\bar{\lambda}_4=\bar{\delta}_1=\bar{\delta}_2=\bar{a}_7=\bar{a}_8=0.01$, $\Delta_1=0.04$, $\Delta_2=\bar{\delta}_3=\bar{a}_2=0.02$, and $\tilde{x}_0=1$. These plots illustrate the differences in the behavior of lower and higher-order moments for different initial states of the internal degree of freedom of the OQBM walker.}
\label{example3}
\end{figure}

Figure~\ref{example3} depicts the $n$-th moments with respect to the imaginary part of the off-diagonal element of the OQBM density matrix $\langle \xi^n \tilde{C}_I (t) \rangle$ (solid curve) and the $n$-th moments with respect to the inverse population in the internal degree of freedom of the open quantum Brownian particle $\langle \xi^n\hat{\sigma}_z (t) \rangle$ (dashed curve) as a function of dimensionless time for various parameters. Figure~\ref{example3}(a) shows the second-order moment ($n=2$) with the internal degree of freedom initialized in a pure state with zero coherences. In contrast, Fig.~\ref{example3}(b) shows the fifteenth-order moment ($n=15$) with the internal degree of freedom initialized in a mixed state with nonzero coherence. As shown in Fig.~\ref{example3}, the system exhibits initial coherences for both low- and high-order moments (e.g., $n=2$ and $n=15$). However, due to decoherence, both $\langle \xi^n \tilde{C}_I (t) \rangle$ and $\langle \xi^n\hat{\sigma}_z (t) \rangle$ decay to zero. We also found that the lower-order moments exhibit larger oscillation amplitudes than higher-order moments, indicating that the system might be experiencing non-Gaussian effects. From the numerical experiments, we found that the higher-order moments (e.g., $n=8,10,15$) have essentially identical dynamics, and that these oscillations and coherences depend strongly on the driving strength $\Omega$, e.g., stronger driving enhances coherences and oscillations, whereas weaker driving suppresses them.

\begin{figure}[htp]
\centering{%
  \includegraphics[width=8.6cm]{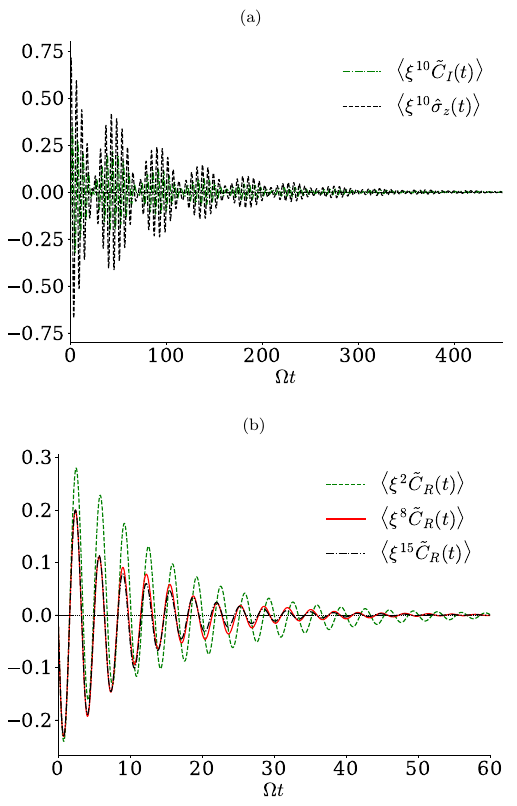}
}
\caption{The time evolution of the $n$-th moments as a function of dimensionless time: the $n$-th moments with respect to the real part, $\langle \xi^n \tilde{C}_R(t) \rangle$,  the imaginary part of the off-diagonal element of the OQBM density matrix, $\langle \xi^n \tilde{C}_I(t) \rangle$, and the $n$-th moment with respect to the inverse population in the internal degree of freedom of the open quantum Brownian particle, $\langle \xi^n \hat{\sigma}_z(t) \rangle$. The initial distribution is given by Eq.~(\ref{iniit}). Subplot (a) corresponds to the tenth-order moment  $n=10$, and the parameters are $\theta=\pi/6$, $\phi=\pi/4$, $\Omega=0.17$, $\bar{\beta}=\Delta_1=\Delta_2=\Delta_3=\Delta_4=\bar{\lambda}_2=\bar{\lambda}_4=\bar{\delta}_1=\bar{\delta}_2=\bar{a}_7=\bar{a}_8=0.01$, $\bar{\lambda}_3=0.05$, $\bar{\delta}_2=\bar{a}_2=0.02$, and $\tilde{x}_0=1$. Subplot (b) corresponds to $n=2, 8, 15$, respectively, as shown in the legend, and the parameters are $\theta=\phi=\pi/2$, $\Omega=0.1$, $\bar{\beta}=0.001$, $\Delta_1=\bar{\delta}_3=0.02$, $\Delta_2=\Delta_3=\Delta_4=\bar{\lambda}_2=\bar{\lambda}_3=\bar{a}_7=0.01$, $\bar{\lambda}_4=0.04$, $\bar{\delta}_1=\bar{a}_8=0.05$, $\bar{\delta}_2=0.03$, $\bar{a}_2=0.008$, and $\tilde{x}_0=1$.}
\label{example4}
\end{figure}

Figure~\ref{example4}(a) shows the dynamics of the tenth-order moment with respect to the imaginary part of the off-diagonal element of the OQBM density matrix, $\langle \xi^{10} \tilde{C}_I(t) \rangle$, and the tenth-order moment with respect to the inverse population in the internal degree of freedom of the open quantum Brownian particle, $\langle \xi^{10} \hat{\sigma}_z(t) \rangle$. In  Fig~\ref{example4}(a), we initialized the internal degree of freedom of the OQBM walker in a mixed qubit state with nonzero coherences. We found beat-like oscillations in both $\langle \xi^{10} \tilde{C}_I(t) \rangle$ and $\langle \xi^{10} \hat{\sigma}_z(t) \rangle$ which indicates interference between the intrinsic coherent dynamics and dissipative coupling to the thermal bath, revealing both the persistence of coherences and possible non-Gaussian features in the distribution. These beat-like oscillations do not depend on the initial state of the internal degree of freedom of the OQBM walker.
We also found that increasing the driving strength $\Omega$ enhances the number and amplitude of oscillations, but the beat structure remains and is gradually damped due to the decoherent interaction with the thermal bath.

Figure~\ref{example4}(b) shows the dynamics of the $n$-th order moments with respect to the real part of the off-diagonal element of the OQBM density matrix, $\langle \xi^n \tilde{C}_R(t) \rangle$. In this example, we initialized the internal degree of freedom of the OQBM walker in a pure state with zero coherences. We found that $\langle \xi^{n=2,8,15} \tilde{C}_R(t) \rangle$ exhibits initial coherences during the evolution, but due to interaction with the thermal bath, $\langle \xi^{n=2,8,15} \tilde{C}_R(t) \rangle$ decays to zero over time. The results are shown up to time $t=60$, beyond which the values remain zero. We also found that the lower-order moments have a greater oscillation amplitude than the higher-order moments. In contrast, higher-order moments ($n\ge 8$) show nearly indistinguishable, rapidly damped dynamics. To investigate if our system deviates from Gaussianity, in the following subsection, we use the dimensionless form of Eq.~(\ref{system_pde}), which we scaled by $\tilde{x}_0$ to derive the equations for the cumulants of the position distribution of the open Brownian walker and analyze the skewness (third cumulant, $\langle x^3 \rangle_c$).

\subsection{Equations for the cumulants generating function}\label{cumulants_c}

\begin{figure}[htp]
\centering{%
  \includegraphics[width=8.6cm]{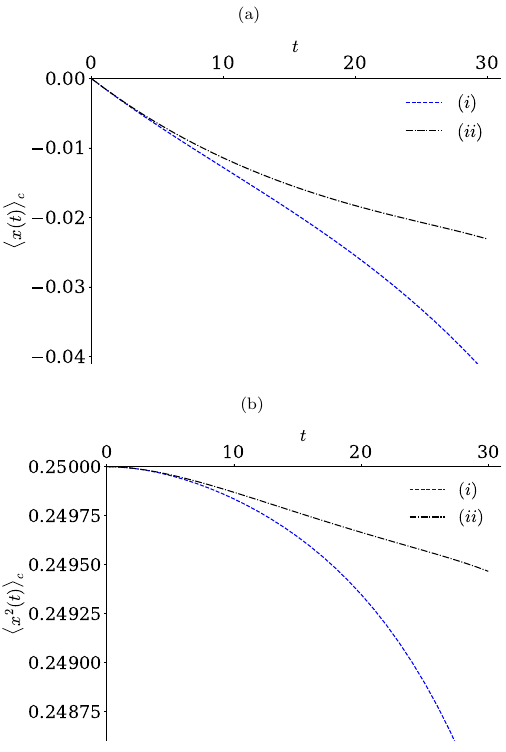}
}
\caption{The time evolution of the first-order cumulant $\langle x \rangle_c$ and the second-order cumulant $\langle x^2 \rangle_c$ as functions of time. Curves ($i$) correspond to $\langle x^n \rangle_c=0$ with $n\geq 3$, while curves ($ii$)  correspond to $\langle x^n \rangle_c=0$ with $n\geq 4$.  The initial distribution is given by Eq.~(\ref{cum1}) with $\theta = \pi/8$ and $\phi = \pi/4$.
Other parameters are $\Omega=0.01, \beta=\bar{\lambda}_2=\Delta_3=0.001, \chi=0.25, \Delta_1=0.05, \bar{\lambda}_3=0.02, \Delta_4=0.1, \bar{a}_2=0.004, \bar{a}_7=0.02, \bar{a}_8=\bar{\delta}_1=0.01,$ and $\bar{\delta}_3=0.002$.}
\label{cum_example1}
\end{figure}

Here, we are interested in investigating the first three cumulants (mean, variance, and skewness) of the position distribution of the OQBM walker. As a first step in our analysis, we define the characteristic function of the scaled reduced density matrix $\tilde{\rho}(\xi,t)$ as
\begin{align}\label{charact}
    \bar{\rho}(k,t) &= \bigl\langle e^{-ik\xi} \bigl\rangle = \int_{-\infty}^{+\infty} d\xi \tilde{\rho} (\xi,t) e^{-ik\xi} \nonumber\\
    &= \sum_{n=0}^\infty \frac{(-ik)^n}{n!}\langle \xi^n\rangle.
\end{align}
\noindent
The expansion of the cumulant generating function~(\ref{charact}) generates the cumulants of the distribution~\cite{kardar2007statistical}, defined as
\begin{align}\label{cum}
    \ln \bar{\rho}(k,t) =\sum_{n=1}^\infty \frac{(-ik)^n}{n!}\langle \xi^n\rangle_c.
\end{align}
\noindent
By substituting Eq.~(\ref{charact}) into the system of partial differential equations~(\ref{system_pde}), we derive the following characteristic system 
\begin{align}\label{system_cum}
&\frac{\partial}{\partial t}\bar{\rho}_+(k,t) = \biggl\{\bar{\lambda}_2\frac{\partial^2}{\partial k^2} -\Delta_1 k \frac{\partial}{\partial k} - \bar{\lambda}_3k^2 -\Delta_1+\lambda_4\biggl\}\bar{\rho}_+\nonumber\\
&-2i\bar{\delta}_1 k \bar{C}_I+2i\bar{a}_2 k \bar{C}_R,\nonumber\\
&\frac{\partial}{\partial t}\bar{\rho}_-(k,t) = \biggl\{\bar{\lambda}_2\frac{\partial^2}{\partial k^2} -\Delta_1 k \frac{\partial}{\partial k} - \bar{\lambda}_3k^2 -\Delta_1+\Delta_2\biggl\}\bar{\rho}_-\nonumber\\
&-4\biggl\{i\bar{a}_7\frac{\partial}{\partial k}-i\bar{\delta}_2 k + \Omega\biggl\}\bar{C}_I+4i\biggl\{\bar{a}_8\frac{\partial}{\partial k}-\bar{\delta}_3k\biggl\}\bar{C}_R\nonumber\\
&-\bar{\beta}\bar{\rho}_+,\nonumber\\
&\frac{\partial}{\partial t}\bar{C}_R(k,t) = \biggl\{\bar{\lambda}_2\frac{\partial^2}{\partial k^2} -\Delta_1 k \frac{\partial}{\partial k} - \bar{\lambda}_3k^2 -\Delta_1+\Delta_3\biggl\}\bar{C}_R\nonumber\\
&-i\biggl\{\bar{a}_8\frac{\partial}{\partial k}-\bar{\delta}_3k\biggl\}\bar{\rho}_-+\frac{i}{2}\bar{a}_2k\bar{\rho}_+,\nonumber\\
&\frac{\partial}{\partial t}\bar{C}_I(k,t) = \biggl\{\bar{\lambda}_2\frac{\partial^2}{\partial k^2} -\Delta_1 k \frac{\partial}{\partial k} - \bar{\lambda}_3k^2 -\Delta_1+\Delta_4\biggl\}\bar{C}_I\nonumber\\
&+\biggl\{i\bar{a}_7\frac{\partial}{\partial k}-i\bar{\delta}_2 k + \Omega\biggl\}\bar{\rho}_--\frac{i}{2}\bar{\delta}_1k\bar{\rho}_+.
\end{align}
\noindent
To simplify the derivation, we cast the above system to
\begin{equation}
    \bar{\rho}_\pm (k,t) = e^{\theta_\pm (k,t)},\hspace{4mm}\text{and}\hspace{4mm}\bar{C}_{R,I} (k,t) = e^{S_{R,I} (k,t)},
\end{equation}
\noindent
and rewrite Eq.~(\ref{system_cum}) as 
\begin{align}\label{rightsystem}
    &\frac{\partial}{\partial t}\theta_+ = \bar{\lambda}_2\biggl( \frac{\partial \theta_+}{\partial k}\biggl)^2+\bar{\lambda}_2\frac{\partial^2 \theta_+}{\partial k^2}-\Delta_1 k\frac{\partial \theta_+}{\partial k}-\bar{\lambda}_3k^2-\Delta_1\nonumber\\
    &+\lambda_4-2i\bar{\delta}_1ke^{(S_I-\theta_+)}+2i\bar{a}_2ke^{(S_R-\theta_+)},\nonumber\\
    &\frac{\partial}{\partial t}\theta_-= \bar{\lambda}_2\biggl( \frac{\partial \theta_-}{\partial k}\biggl)^2+\bar{\lambda}_2\frac{\partial^2 \theta_-}{\partial k^2}-\Delta_1 k\frac{\partial \theta_-}{\partial k}-\bar{\lambda}_3k^2-\Delta_1\nonumber\\
    &+\Delta_4-4e^{(S_I-\theta_-)}\biggl\{i\bar{a}_7\frac{\partial S_I}{\partial k}-i\bar{\delta}_2 k +\Omega\biggl\}-\bar{\beta}e^{(\theta_+-\theta_-)}\nonumber\\
    &+4ie^{(S_R-\theta_-)}\biggl\{\bar{a}_8\frac{\partial S_R}{\partial k}-\bar{\delta}_3 k\biggl\},
    \nonumber\\
    &\frac{\partial}{\partial t}S_R= \bar{\lambda}_2\biggl( \frac{\partial S_R}{\partial k}\biggl)^2+\bar{\lambda}_2\frac{\partial^2 S_R}{\partial k^2}-\Delta_1 k\frac{\partial S_R}{\partial k}-\bar{\lambda}_3k^2-\Delta_1\nonumber\\
    &+\Delta_3-ie^{(\theta_--S_R)}\biggl\{\bar{a}_8\frac{\partial \theta_-}{\partial k}-\bar{\delta}_3 k\biggl\}-\frac{i}{2}\bar{a}_2ke^{(\theta_+-S_R)},
     \nonumber\\
    &\frac{\partial}{\partial t}S_I= \bar{\lambda}_2\biggl( \frac{\partial S_I}{\partial k}\biggl)^2+\bar{\lambda}_2\frac{\partial^2 S_I}{\partial k^2}-\Delta_1 k\frac{\partial S_I}{\partial k}-\bar{\lambda}_3k^2-\Delta_1\nonumber\\
    &+\Delta_4+e^{(\theta_--S_I)}\biggl\{i\bar{a}_2\frac{\partial \theta_-}{\partial k}-i\bar{\delta}_2 k+\Omega\biggl\}-\frac{i}{2}\bar{\delta}_1ke^{(\theta_+-S_I)}.
\end{align}
\noindent
Using the following relations
\begin{align}
    &\theta_+ = \sum_{n=0}^\infty \frac{(-ik)^n}{n!}\langle x^n \rangle_c,\hspace{2mm} S_R = \sum_{n=0}^\infty \frac{(-ik)^n}{n!}\langle r^n \rangle_c\nonumber\\
    &\theta_- = \sum_{n=0}^\infty \frac{(-ik)^n}{n!}\langle z^n \rangle_c,\hspace{2mm} S_I = \sum_{n=0}^\infty \frac{(-ik)^n}{n!}\langle i^n \rangle_c,
\end{align}
\noindent
one can straightforwardly show that their derivatives and exponential expansions are
\begin{align}\label{deriva}
    &\frac{\partial}{\partial t}\theta_+ = \sum_{n=0}^\infty \frac{(-ik)^n}{n!}\frac{\partial}{\partial t}\langle x^n \rangle_c, \nonumber\\
    &\frac{\partial}{\partial k}\theta_+ =-i \sum_{n=0}^\infty \frac{(-ik)^n}{n!}\langle x^{n+1} \rangle_c,\nonumber\\
    &\biggl(\frac{\partial}{\partial k}\theta_+ \biggl)^2=-\sum_{m,n=0}^\infty \frac{(-ik)^{n+m}}{n!m!}\langle x^{n+1} \rangle_c\langle x^{m+1} \rangle_c,\nonumber\\
    &\frac{\partial^2}{\partial k^2}\theta_+ =- \sum_{n=0}^\infty \frac{(-ik)^n}{n!}\langle x^{n+2} \rangle_c,\nonumber\\
     &e^{S_I-\theta_+} = \sum_{n=0}^\infty \sum_{m_1+\cdots m_n=0 }^\infty \frac{1}{n!}\frac{(-ik)^{m_1+\cdots m_n}}{m_1!m_2!\cdots m_n!}\Delta_{m_1}\nonumber\\
     &\times \Delta_{m_2}\cdots \Delta_{m_n},
\end{align}
\noindent
where, 
\begin{equation}
    \Delta_{m_n} = \langle i^{m_n} \rangle -\langle x^{m_n} \rangle.
\end{equation}
\noindent
To analyze this system~(\ref{system_cum}), we perform a cumulant truncation analysis. As a first example, we assume that the system can be approximated as a Gaussian process.
Since Gaussian processes have vanishing higher-order cumulants, i.e., $\langle x^n \rangle_c = 0$ for $n\geq 3$, the cumulants for this system~(\ref{system_cum}) can be derived explicitly. For the zeroth order term $(-ik)^0$, we obtain the following expression
\begin{equation}\label{zeroth_0}
  0 = \bar{\lambda}_2\langle x \rangle_c^2 +\bar{\lambda}_2\langle x^2 \rangle_c +\Delta_1-\lambda_4.
  \end{equation}
\noindent
By substituting the corresponding parameters, Eq.~(\ref{zeroth_0}) can be cast into the following form
\begin{equation}\label{zeroth}
  \langle x \rangle_c^2 +  \langle x^2 \rangle_c = \langle x^2 \rangle = \frac{\hbar}{2m\omega \tilde{x}_0^2}\bigl[2n(\omega)+1\bigl].
\end{equation}
\noindent
Equation~(\ref{zeroth}) shows that the second moment $\langle x^2 \rangle$ of the position distribution of the OQBM walker originates from both the quantum zero-point energy and the thermal occupation $n(\omega)$, which means that $\langle x^2 \rangle$ does not vanish in the ground state due to vacuum fluctuations.
Here, the second-order cumulant $\langle x^2 \rangle_c$ describes the intrinsic quantum fluctuations, while the first-order cumulant  $\langle x \rangle_c$ represents the mean position of the OQBM walker. 
From Eq.~(\ref{zeroth}), for simplicity, we write $\langle x^2 \rangle_c$ as
\begin{equation}\label{zeroth2}
   \langle x^2 \rangle_c = \chi -\langle x\rangle_c^2,
\end{equation}
\noindent
where $\chi=\frac{\hbar}{2m\omega \tilde{x}_0^2}\bigl[2n(\omega)+1\bigl]$. 
Equation~(\ref{zeroth2}) allows us to compute the variance $\langle x^2 \rangle_c$ directly from the mean $\langle x \rangle_c$. 
The remaining cumulant equations, such as those for $\langle x \rangle_c $ and $\langle z^0\rangle_c $, are given by
\begin{align}\label{cum_eq1}
\frac{d}{dt}\langle x& \rangle_c =-2\bar{\lambda}_2\chi\langle x \rangle_c +2\bar{\lambda}_2\langle x \rangle_c^3-\Delta_1\langle x \rangle_c  +2\bar{\delta}_1e^{\langle i^0\rangle- 1}\nonumber\\
&-2\bar{a}_2e^{\langle r^0\rangle - 1},\nonumber\\
\frac{d }{d t}\langle z^0 &\rangle_c =-\bar{\lambda}_2 \langle z \rangle_c^2-\bar{\lambda}_2 \langle z^2 \rangle_c-\Delta_1+\Delta_4-4\bigl\{\bar{a}_7\langle i \rangle_c\nonumber\\
&+\Omega\bigl\}e^{\langle i^0\rangle_c-\langle z^0\rangle_c}-\bar{\beta}e^{1-\langle z^0\rangle_c}+4\bar{a}_8\langle r \rangle_c e^{\langle r^0\rangle_c-\langle z^0 \rangle_c}.
\end{align}
\noindent
The rest of the cumulant equations are given explicitly in the Appendix~\ref{cumulants} (see Table~\ref{tab:appendix_eqns1}). The system described by Eq.~(\ref{cum_eq1}) is then solved numerically for a Gaussian initial distribution of the form:
\begin{equation}\label{cum1}
  \tilde{\rho}(\xi,0) = \frac{1}{2\sqrt{\pi}}e^{-\xi^2}\otimes\begin{pmatrix}
2\cos^2\theta & \sin 2\theta e^{-i\phi} \\
\sin 2\theta e^{i\phi} & 2\sin^2\theta
\end{pmatrix}.
\end{equation}
\noindent
To derive the initial conditions for this system~(\ref{cum_eq1}), we use
\begin{equation}\label{initi}
   f(\hat{X}) = \ln\biggl[ \int_{-\infty}^{+\infty}  d\xi e^{-ik\xi} \tr\bigl\{\hat{X}\tilde{\rho}(\xi,0)\bigl\}\biggl],
\end{equation}
\noindent
where $f\bigl\{\hat{I},\frac{1}{2}\hat{\sigma}_x,-\frac{1}{2}\hat{\sigma}_y\bigl\}=\bigl\{\theta_+, \theta_-, S_R, S_I\bigl\}(k,0)$. From Eqs.~(\ref{cum1})-(\ref{initi}), it is straightforward to show that the initial conditions are 
\begin{align}
 \theta_+(k,0) &= \frac{1}{4}(-ik)^2-\frac{1}{2}\ln\pi,\nonumber\\
 \theta_-(k,0) &= \frac{1}{4}(-ik)^2+\ln\biggl[\frac{\cos2\theta}{\sqrt{\pi}}\biggl],\nonumber\\
  S_R(k,0) &= \frac{1}{4}(-ik)^2+\ln\biggl[\frac{\sin2\theta\cos\phi}{2\sqrt{\pi}}\biggl],\nonumber\\
   S_I(k,0) &= \frac{1}{4}(-ik)^2+\ln\biggl[\frac{e^{i\pi}\sin2\theta\cos\phi}{2\sqrt{\pi}}\biggl],
\end{align}
\noindent
meaning that at time $t=0$, we have
\begin{align}
    \langle x^0 (0)\rangle_c &= -\frac{1}{2}\ln\pi, \hspace{1mm} \langle x (0)\rangle_c =0, \hspace{1mm}  \langle z^0 (0)\rangle_c =\ln\biggl[\frac{\cos2\theta}{\sqrt{\pi}}\biggl],\nonumber\\
    \langle r^0 (0)\rangle_c &=\ln\biggl[\frac{\sin2\theta\cos\phi}{2\sqrt{\pi}}\biggl],\hspace{1mm} \langle i^0(0) \rangle_c =\ln\biggl[\frac{e^{i\pi}\sin2\theta\sin\phi}{2\sqrt{\pi}}\biggl],\nonumber\\
     \langle x^2(0) \rangle_c&=\langle z^2 (0)\rangle_c=\langle i^2 (0)\rangle_c=\langle r^2(0) \rangle_c=\chi=\frac{1}{4}.
\end{align}
\noindent
Figure~\ref{cum_example1} depicts (a) the first-order cumulant $\langle x \rangle_c$ and (b) the second-order cumulant $\langle x^2 \rangle_c$ of the OQBM walker as a function of time for various parameters.
For curves ($i$) in Fig.~\ref{cum_example1}, we assume Gaussian closure ($\langle x^n\rangle_c=0$ for $n\geq 3$). The strong nonlinearities in the cumulant equations lead to numerical instability for $t \geq 30$, limiting the displayed time range.
Figure~\ref{cum_example1}(a), curve ($i$), shows that the average position $\langle x \rangle_c$ of the OQBM walker starts at zero and slowly drifts roughly linearly toward negative values over time. This drift indicates a shifted displacement induced by the thermal bath.
Figure~\ref{cum_example1}(b), curve($i$), shows that the second cumulant starts at $\langle x^2 (0) \rangle_c = 0.25$ and then decreases smoothly as a function of time. This decrease reflects bath-induced localization of the walker's position. In contrast to the traditional classical Brownian motion, where the variance grows due to diffusion, the OQBM walker undergoes bath-induced localization. 



As a second example, we assume that the system can be approximated as a Gaussian process, and derive the cumulants for the case where $\langle x^n \rangle_c = 0$ for $n\geq 4$. Proceeding as previously, one can show that for the zeroth-order term $(-ik)^0$, we obtain
\begin{equation}
  0 = \bar{\lambda}_2\langle x \rangle_c^2 +\bar{\lambda}_2\langle x^2 \rangle_c +\Delta_1-\lambda_4.
  \end{equation}
\noindent
The above equation is the same as the one presented in Eq.~(\ref{zeroth_0}). 
The first cumulant $\langle x \rangle_c$ and the third cumulant $\langle x^3 \rangle_c$ of the position of the OQBM walker are given explicitly by the following nonlinear expressions, respectively: 
\begin{align}\label{4th_zerocumulants}
\frac{d}{dt}\langle x \rangle_c &=-\bar{\lambda}_2\langle x^3 \rangle_c-(2\bar{\lambda}_2\chi+\Delta_1)\langle x \rangle_c +2\bar{\lambda}_2\langle x \rangle_c^3\nonumber\\
&+2\bar{\delta}_1e^{\langle i^0\rangle_c- 1}-2\bar{a}_2e^{\langle r^0\rangle_c - 1},\nonumber\\
\frac{d }{d t}\langle x^3 \rangle_c &=-3\Delta_1\langle x^3 \rangle_c-6\bar{\lambda}_2\chi\langle x^3\rangle_c+6\bar{\lambda}_2\langle x\rangle_c^2\langle x^3\rangle_c\nonumber\\
&-6\bar{a}_2e^{\langle r^0\rangle_c-1}\bigl\{\langle r\rangle_c^2+\langle r^2\rangle_c-2\langle r\rangle_c\langle x\rangle_c+2\langle x\rangle_c^2\nonumber\\
&-\chi\bigl\}+6\bar{\delta}_1e^{\langle i^0\rangle_c-1}\bigl\{\langle i\rangle_c^2+\langle i^2\rangle_c-2\langle i\rangle_c\langle x\rangle_c\nonumber\\
&+2\langle x\rangle_c^2-\chi\bigl\}.
\end{align}
\noindent
Other cumulants are provided in the Appendix~\ref{cumulants} (see Table~\ref{tab:appendix_eqns2}).
The above system~(\ref{4th_zerocumulants}) is then solved numerically for various parameters.
Figure~\ref{cum_example1} shows (a) curve$(ii)$, the first-order cumulant $\langle x \rangle_c$, and (b) curve$(ii)$, the second-order cumulant $\langle x^2 \rangle_c$ as a function of time. 
We found that even when the higher-order cumulants are set to zero starting from the fourth order, both $\langle x \rangle_c$  and $\langle x^2 \rangle_c$ exhibit similar dynamics up to approximately $t\approx 10$ to those observed when the truncation begins at the third order. For time $t<10$, the system behaves as a Gaussian process. However, for $t > 10$, as time progresses, the cumulants deviate significantly, indicating that the third-order cumulants play a crucial role and warrant further investigation. In this parameter regime, the system clearly exhibits distinctly non-Gaussian behavior.

\begin{figure}[htbp]
\centering{%
  \includegraphics[width=8.6cm]{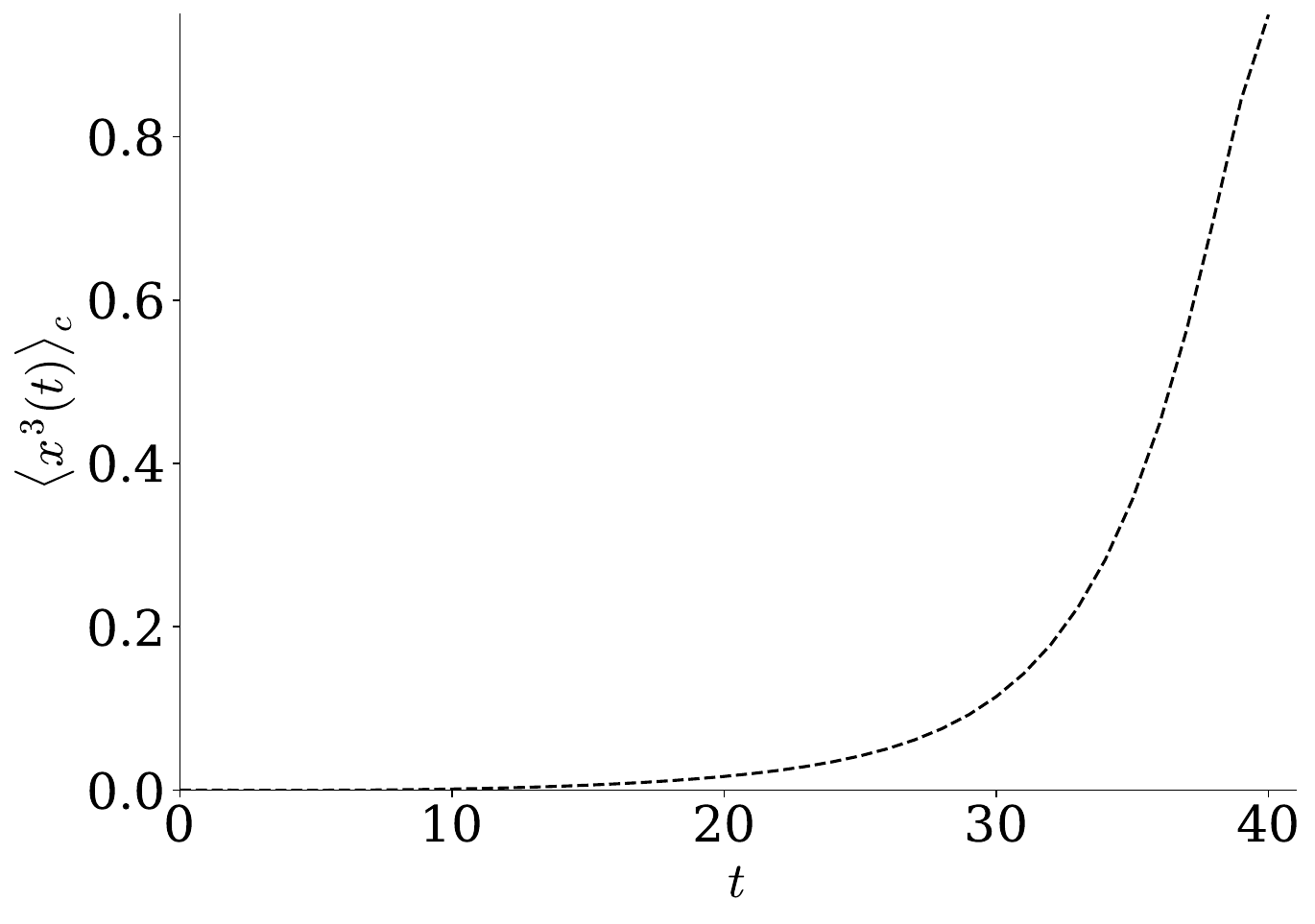}
}
\caption{The time evolution of the third-order cumulant $\langle x^3 \rangle_c$ as functions of time. The initial distribution is given by Eq.~(\ref{cum1}) with $\theta = \pi/8$ and $\phi = \pi/4$.
Other parameters are $
\Omega=0.01, \beta=\bar{\lambda}_2=\Delta_3=0.001, \chi=0.25, \Delta_1=0.05, \bar{\lambda}_3=0.02, \Delta_4=0.1, \bar{a}_2=0.004, \bar{a}_7=0.02, \bar{a}_8=\bar{\delta}_1=0.01,$
 and $\bar{\delta}_3=0.002$.}
\label{example_cum4}
\end{figure}

Figure~\ref{example_cum4} shows the third-order cumulant $\langle x^3 \rangle_c$ (skewness) as a function of time.  For this OQBM walker, the skewness remains nonzero and increases exponentially with time, indicating that the intrinsic generator of the dynamics is non-Gaussian.

\section{Summary and conclusion}\label{conclussion}

In this paper, we presented the generic case of the microscopic derivation of a completely positive quantum master equation that describes Open Quantum Brownian Motion. We began with a microscopic Hamiltonian, which comprises the Hamiltonian of a weakly driven open Brownian particle in a harmonic potential, the Hamiltonian of a thermal bath, and the system-bath interaction Hamiltonian. We applied the Born-Markov approximation and the rotating wave approximation to the system-bath interaction Hamiltonian to obtain the reduced quantum master equation for the system density matrix. The derived quantum master equation has the same operatorial structure as that of the Lindblad master equation~\cite{gorini1976completely,lindblad1976generators} and therefore guarantees completely positive dynamics. The resulting master equation was written in a generic non-diagonal position representation, and we then performed adiabatic elimination of the fast variables to obtain the master equation that defines OQBM.

We illustrated the derived dynamics using both Gaussian and non-Gaussian initial spatial distributions, with the internal degree of freedom initially in a pure state. We found that the position probability distribution of finding the open quantum Brownian particle at certain positions for different moments of time is given by a few Gaussian distributions, even for explicitly non-Gaussian initial states. 
This behavior arises from the internal degree of freedom of the OQBM walker, which transfers its coherent component to the spatial part, resulting in multiple Gaussian profiles propagating at different spreading rates. We then plotted the time-dependent variance $\sigma^2(t)$  in the position of the OQBM walker and observed a Gaussian-like dynamics. 
We also analyzed the off-diagonal coherence $(C_I(t)=\tr_x\bigl[C_I(x,t)\bigl])$ and the population inversion $\langle \hat{\sigma}_z (t) \rangle$. Both quantities exhibit initial coherence before decaying to zero due to interaction with the bath.

With the help of the obtained quantum master equation for the OQBM, we derived equations for the $n$-th moments and the cumulants of the position distribution of the OQBM walker. We numerically solved these equations for Gaussian initial distributions for various parameter regimes. We found that the higher-order moments exhibit identical dynamics, and the coherences depend strongly on the driving strength of the internal degree of freedom of the OQBM walker. Finally, a cumulant truncation analysis reveals a nonzero third-order cumulant $\langle x^3 \rangle_c$, confirming the non-Gaussian character of the dynamics.

In conclusion, we have provided a first-principles derivation of a completely positive master equation for OQBM in a harmonic potential, resolving the positivity issues of earlier CL-based models and explicitly revealing its hybrid quantum-classical structure.
In future work, it will be interesting to develop microscopic Hamiltonians, which include nonlinear system-bath coupling, but with the assumption of factorized initial conditions in the derivation of OQBM, and derive OQBM for cases where the open Brownian particle is confined within an anharmonic potential, and explore the extension to higher dimensions in internal degree of freedom and position space.
Since the OQBM master equation is a typical example of hybrid quantum-classical master equations, which find application in various fields, including the generalization of gravity~\cite{layton2024healthier,layton2024classical,oppenheim2023postquantum,oppenheim2022two,halliwell1998effective,tilloy2024general}. Deriving OQBM in such settings would be an interesting direction for future work.

\begin{acknowledgments}

This work is based upon research supported by the National Research Foundation (NRF) of the Republic of South Africa. AZ acknowledges support in part by the NRF of South Africa (Grant No. 129457). 
\end{acknowledgments}

\appendix
\section{Derivation of Eq.~(\ref{oqbm_me})}\label{appendixa}
Here, we present the adiabatic elimination of fast variables method~\cite{smoluchowski1916brownsche,van1985elimination,gardiner1985handbook,kramers1940brownian}. As the first step, we write Eq.~(\ref{uv_eq}) as
\begin{align}\label{weqfin3a}
   &\frac{\partial}{\partial t}\rho(u,v) =\Bigl(\bar{\alpha}_1 \hat{L}_1 +\hat{L}_2\Bigl)\rho+\biggl(\bar{\alpha}_2\hbar^2\frac{\partial^2}{\partial u^2}+2\bar{\alpha}_3\hbar\nonumber\\
   &+2\bar{\alpha}_3\hbar u\frac{\partial}{\partial u}\biggl)\rho+\biggl(\frac{\partial}{\partial u} \hat{m}_1+\frac{\partial}{\partial v} \hat{m}_2+ u \hat{m}_3
   +v \hat{m}_4\biggl)\rho \nonumber\\
   &+\pazocal{L}_{\mathrm{2LS}} \rho,
\end{align}
\noindent
where $\hat{L}_1$ and $\hat{L}_2$, are defined as 
\begin{subequations}
  \begin{align}
     \hat{L}_1 & = -v^2-\gamma v \frac{\partial}{\partial v}, \label{eq_1q}\\
    \hat{L}_2 & = \frac{i\hbar}{m}\frac{\partial^2}{\partial v \partial u}-\frac{i}{\hbar}m\omega^2uv, \label{eq_2}
\end{align}
\end{subequations}
\noindent
with $\gamma = 2\hbar \bar{\alpha}_3 /\bar{\alpha}_1 $. We assume that $(m\omega)^2\sim k_BT$ and we also assume that $\bar{\alpha}_1$ is much greater than all the system parameters, which is the time scale for the fast variable $v$. Our goal is to derive a distribution function $\bar{\rho}(u)$ for the slow variable $u$ of the system, defined as
\begin{equation}
\bar{\rho}(u) = \int_{-\infty}^{+\infty} dv \rho(u,v),
\end{equation}
\noindent
which will be valid in the limit when $\bar{\alpha}_1$ becomes very large.
We obtain the stationary distribution of Eq.~(\ref{weqfin3a}) by solving the differential operator~(\ref{eq_1q}), which takes the form:
\begin{equation}\label{a111}
     \frac{\partial}{\partial t} \rho_s = \hat{L}_1\rho_s=-\biggl(v^2\rho_s+\gamma v\frac{\partial }{\partial v}\rho_s \biggl)=0.
\end{equation}
\noindent
By solving Eq.~(\ref{a111}), we obtain the normalized stationary solution $\rho_s(v)$, which is given by
\begin{equation}
    \rho_s(v)=(2\pi\gamma)^{-1/2}\exp(-v^2/2\gamma).
\end{equation}
\noindent
We now introduce a projection operator $\mathcal{P}$, which reads
\begin{equation}\label{aa1}
\mathcal{P}f(u,v) = \rho_s(v)\int dv' f(u,v'),
\end{equation}
\noindent
where $f(u,v)$ is an arbitrary function.
Applying $\mathcal{P}$ to $\rho(u,v)$ yields
\begin{equation}\label{pw}
\mathcal{P}\rho(u,v) = \rho_s(v)\bar{\rho}(u).
\end{equation}
\noindent
Equation~(\ref{aa1}) can now be written as
\begin{equation}\label{aa12a}
g(u,v) = \rho_s(v)\hat{g}(u),
\end{equation}
\noindent
where $g(u,v)$ is an arbitrary function.
Functions of type~(\ref{aa12a}) are all solutions of $\hat{L}_1 g = 0$,
that is, the space that $\mathcal{P}$ projects onto is the null space of $\hat{L}_1$. 
Other properties of $\mathcal{P}$, which can be verified through a straightforward brute-force calculation, are:
\begin{subequations}
\begin{align}
&\mathcal{P}^2=\mathcal{P},\label{eqwa}\\
&\hat{L}_1\mathcal{P} = \mathcal{P}\hat{L}_1=0,\label{eqwb}\\
&\mathcal{P}\hat{L}_2\mathcal{P}=0,\label{eqwc}\\
&\mathcal{P} = \lim_{t\to\infty}\bigl[\exp\bigl(\hat{L}_1 t\bigl)\bigl]\label{eqw}.
\end{align}
\end{subequations}
\noindent
Properties~(\ref{eqwa}),~(\ref{eqwb}), and~(\ref{eqwc}) can easily be verified.
To verify~(\ref{eqw}), we can expand any function of $u$ and $v$ in eigenfunctions $P_\lambda(v)$ of $\hat{L}_1$ as
\begin{subequations}
\begin{align}
f(u,v) =\ & \sum_{\lambda} A_\lambda (u) P_\lambda(v), \\
\text{where,} \hspace{4mm} A_\lambda(u) = \ & \int dp Q_\lambda (v) f(u,v).
\end{align}
\end{subequations}
\noindent
Then, the long-time limit of~(\ref{eqw}) can be written as
\begin{align}
\lim_{t\to\infty}\bigl[\exp\bigl(\hat{L}_1 t\bigl) &f(u,v)\bigl]  =\sum_{\lambda} A_\lambda (u)  \lim_{t\to\infty}e^{-\lambda t} P_\lambda(v)\nonumber\\
 &=P_0(v)\int dv Q_0(v) f(u,v),
\end{align}
\noindent
where $P_0(v) =\rho_s(v)$ and $Q_0(v)=1$.
By using the property~(\ref{eqwc}),
it is straightforward to verify that for this process $v\exp(-v^2/2\gamma)\propto P_1(v)$, and  $\mathcal{P}P_1(v)=0$.
The properties~(\ref{eqwa})-(\ref{eqw}) will be repeatedly used throughout this paper.

The next step is to define another operator  $\mathcal{Q}$, i.e., $\mathcal{Q}=1-\mathcal{P}$, where the operators $\mathcal{P}$ and $\mathcal{Q}$ select the relevant part and the irrelevant part of $\rho(u,v)$. The standard properties of projectors $\mathcal{Q}^2=\mathcal{Q}$ and $\mathcal{P}\mathcal{Q}=\mathcal{Q}\mathcal{P}=0$, applies. Following the projection operator method, one writes
\begin{subequations}
\begin{align}
\bar{v}  &= \mathcal{P}\rho, \\
\bar{w}&= \mathcal{Q}\rho=(1-\mathcal{P})\rho,
\end{align}
\end{subequations}
\noindent
where the function $\bar{v}= P_0(v) \bar{\rho}(u)$ varies on the slow time scale, and $\bar{w}$ varies on the fast time scale. As a result, the distribution $\rho(u,v)$ can now be decomposed into two parts
\begin{equation}\label{weqfin4}
    \rho(u,v) = \bar{v}+\bar{w}.
\end{equation}
\noindent
Applying the projector $\mathcal{P}$ and $(1-\mathcal{P})$ to Eq.~(\ref{weqfin3a}), yields
\begin{subequations}
\begin{align}
  & \frac{\partial \bar{v}}{\partial t} =  \mathcal{P}\bigl(\bar{\alpha}_1 \hat{L}_1 +\hat{L}_2\bigl)\rho+\mathcal{P}\biggl(\frac{\partial}{\partial u} \hat{m}_1+\frac{\partial}{\partial v} \hat{m}_2+ u \hat{m}_3\nonumber\\
   &+v \hat{m}_4\biggl)\rho +\mathcal{P}\biggl(\bar{\alpha}_2\hbar^2\frac{\partial^2}{\partial u^2}+2\hbar\bar{\alpha}_3+2\hbar\bar{\alpha}_3u\frac{\partial}{\partial u}\biggl)\rho\nonumber\\
&+\mathcal{P}\pazocal{L}_{\mathrm{2LS}} \rho,\\
   &\frac{\partial \bar{w}}{\partial t} = \mathcal{Q}\bigl(\bar{\alpha}_1 \hat{L}_1 +\hat{L}_2\bigl)\rho+\mathcal{Q}\biggl(\frac{\partial}{\partial u} \hat{m}_1+\frac{\partial}{\partial v} \hat{m}_2+ u \hat{m}_3\nonumber\\
   &+v \hat{m}_4\biggl)\rho +\mathcal{Q}\biggl(\bar{\alpha}_2\hbar^2\frac{\partial^2}{\partial u^2}+2\hbar\bar{\alpha}_3+2\hbar\bar{\alpha}_3u\frac{\partial}{\partial u}\biggl)\rho\nonumber\\
&+\mathcal{Q}\pazocal{L}_{\mathrm{2LS}} \rho.
\end{align}
\end{subequations}

\noindent
After some algebra and utilizing property~(\ref{eqwc}) and $\mathcal{P}\pazocal{L}_{\mathrm{2LS}}\mathcal{P}=\pazocal{L}_{\mathrm{2LS}}\mathcal{P}=\mathcal{P}\pazocal{L}_{\mathrm{2LS}}$, we obtain
\begin{subequations}\label{abc11}
\begin{align}
 \frac{\partial \bar{v}}{\partial t} & = \mathcal{P}\hat{L}_2\bar{w} + \hat{m}_1 \frac{\partial \bar{v}}{\partial u} +u\hat{m}_3 \bar{v}+\bar{\alpha}_2\hbar^2\frac{\partial^2\bar{v}}{\partial u^2} +2\bar{\alpha}_3\hbar\bar{v}\nonumber\\
 &+2\bar{\alpha}_3\hbar u \frac{\partial \bar{v}}{\partial u}+\pazocal{L}_{\mathrm{2LS}}\bar{v},\\
 \frac{\partial \bar{w}}{\partial t} & =  \bar{\alpha}_1 \hat{L}_1\bar{w}+(1-\mathcal{P})\hat{L}_2\bar{w} +\hat{L}_2\bar{v}+\hat{m}_1\frac{\partial \bar{w}}{\partial u} + \hat{m}_2\frac{\partial \bar{v}}{\partial v}\nonumber\\
 &+ \hat{m}_2\frac{\partial \bar{w}}{\partial v}+u\hat{m}_3\bar{w}+v\hat{m}_4\bar{v}+v\hat{m}_4\bar{w}+\bar{\alpha}_2\hbar^2\frac{\partial^2\bar{w}}{\partial u^2}\nonumber\\
&+2\hbar\bar{\alpha}_3\bar{w}+2\hbar\bar{\alpha}_3u\frac{\partial\bar{w}}{\partial u}+\pazocal{L}_{\mathrm{2LS}}\bar{w}.
\end{align}
\end{subequations}

We solve the above equations~(\ref{abc11}) using the Laplace transform in the next subsection.

\subsection{Solution via Laplace Transform}

We adopt the following notation for the Laplace transform
\begin{equation}\label{laplace}
\tilde{h}(s) = \int_0^\infty dt e^{-st} h(t),  
\end{equation}
\noindent
where $h(t)$ is an arbitrary function of time. Applying Eq.~(\ref{laplace}) into Eq.~(\ref{abc11}) yields
\begin{subequations}
\begin{align}
&s\tilde{v}(s)  =  \mathcal{P}\hat{L}_2\tilde{w}(s)+ \hat{m}_1 \frac{\partial \tilde{v}(s)}{\partial u} +u\hat{m}_3 \tilde{v}(s) +\bar{\alpha}_2\hbar^2\frac{\partial^2 \tilde{v}(s)}{\partial u^2}\nonumber\\
&+2\hbar\bar{\alpha}_3\tilde{v}(s)+2\hbar\bar{\alpha}_3u\frac{\partial \tilde{v}(s)}{du}+\pazocal{L}_{\mathrm{2LS}}\tilde{v}(s)+v(0), \label{cn1}\\
&s\tilde{w}(s) = \bar{\alpha}_1 \hat{L}_1\tilde{w}(s)+(1-\mathcal{P})\hat{L}_2\tilde{w}(s) +\hat{L}_2\tilde{v}(s)\nonumber\\
&+\hat{m}_1\frac{\partial \tilde{w}(s)}{\partial u} + \hat{m}_2\frac{\partial \tilde{v}(s)}{\partial v}+ \hat{m}_2\frac{\partial \tilde{w}(s)}{\partial v}+u\hat{m}_3\tilde{w}(s)\nonumber\\
&+v\hat{m}_4\tilde{v}(s)+w(0)+v\hat{m}_4\tilde{w}(s)+\bar{\alpha}_2\hbar^2\frac{\partial^2 \tilde{w}(s)}{\partial u^2}\nonumber\\
&+2\hbar\bar{\alpha}_3\tilde{w}(s)+2\hbar\bar{\alpha}_3u\frac{\partial \tilde{w}(s)}{\partial u}+\pazocal{L}_{\mathrm{2LS}}\tilde{w}(s).\label{abcs}
\end{align}
\end{subequations}
\noindent
We then assume that $w(0)=0$, which means that the initial distribution is of the form 
\begin{equation}\label{init_v}
    \rho(u,v,t=0) = (2\pi\gamma)^{-1/2}\exp(-v^2/2\gamma)\bar{\rho}(u).
\end{equation}
\noindent
The above~(\ref{init_v}) satisfies the condition of initial thermalization of variable $v$. We then solve Eq.~(\ref{abcs}) for $\tilde{w}(s)$ to obtain
\begin{align}\label{abcs1}
&\tilde{w}(s)=\ \bigl[s-\bar{\alpha}_1\hat{L}_1 -(1-\mathcal{P})\hat{L}_2-\hat{m}_1\frac{\partial}{\partial u}-\hat{m}_2\frac{\partial}{\partial v}-u\hat{m}_3\nonumber\\
&-v\hat{m}_4-\bar{\alpha}_2\hbar^2\frac{\partial^2}{\partial u^2}-2\hbar\bar{\alpha}_3-2\hbar\bar{\alpha}_3u\frac{\partial}{\partial u}-\pazocal{L}_{\mathrm{2LS}}\bigl]^{-1}\nonumber\\
&\times \biggl(\hat{L}_2+v\hat{m}_4+\hat{m}_2\frac{\partial}{\partial v}\biggl)\tilde{v}(s).
\end{align}
\noindent
Substituting Eq.~(\ref{abcs1}) into Eq.~(\ref{cn1}), yields
\begin{widetext}
\begin{align}\label{cn1q}
&s\tilde{v}(s)-v(0)= \mathcal{P}\hat{L}_2\Bigl[s-\bar{\alpha}_1 \hat{L}_1 -(1-\mathcal{P})\hat{L}_2-\hat{m}_1\frac{\partial}{\partial u}-\hat{m}_2\frac{\partial}{\partial v}-u\hat{m}_3-v\hat{m}_4-\bar{\alpha}_2\hbar^2\frac{\partial^2}{\partial u^2}-2\hbar\bar{\alpha}_3-2\hbar\bar{\alpha}_3u\frac{\partial}{\partial u}\nonumber\\
&-\pazocal{L}_{\mathrm{2LS}}\Bigl]^{-1} \biggl(\hat{L}_2+v\hat{m}_4+\hat{m}_2\frac{\partial}{\partial v}\biggl)\tilde{v}(s)+\hat{m}_1\frac{\partial \tilde{v}(s)}{\partial u}+u\hat{m}_3\tilde{v}(s)+\bar{\alpha}_2\hbar^2\frac{\partial^2 \tilde{v}(s)}{\partial u^2}+2\hbar\bar{\alpha}_3\tilde{v}(s)+2\hbar\bar{\alpha}_3u\frac{\partial \tilde{v}(s)}{\partial u}+\pazocal{L}_{\mathrm{2LS}}\tilde{v}(s).
\end{align}
\end{widetext}
\noindent
Here, we have a partial solution to the problem. For any finite $s$, we take the limit of large $\bar{\alpha}_1$, which yields
\begin{align}\label{fine_eqa}
&s\tilde{v}(s) - v(0)\approx -(\bar{\alpha}_1)^{-1}\mathcal{P}\hat{L}_2\hat{L}_1^{-1}\biggl(\hat{L}_2+v\hat{m}_4+\hat{m}_2\frac{\partial}{\partial v}\biggl)\nonumber\\
&\times\tilde{v}(s)+\hat{m}_1\frac{\partial \tilde{v}(s)}{\partial u}+u\hat{m}_3\tilde{v}(s)+\bar{\alpha}_2\hbar^2\frac{\partial^2 \tilde{v}(s)}{\partial u^2}+2\hbar\bar{\alpha}_3\tilde{v}(s)\nonumber\\
&+2\hbar\bar{\alpha}_3u\frac{\partial \tilde{v}(s)}{\partial u}+\pazocal{L}_{\mathrm{2LS}}\tilde{v}(s).
\end{align}
\noindent
We then return to the time domain to find
\begin{align}\label{felimp}
 &\frac{\partial \bar{v}}{\partial t}= -(\bar{\alpha}_1)^{-1}\mathcal{P}\hat{L}_2\hat{L}_1^{-1}\hat{L}_2\bar{v}\nonumber\\
&-(\bar{\alpha}_1)^{-1}\mathcal{P}\hat{L}_2\hat{L}_1^{-1}\biggl(v\hat{m}_4+\hat{m}_2\frac{\partial}{\partial v}\biggl)\bar{v}+\hat{m}_1\frac{\partial \bar{v}}{\partial u}+u\hat{m}_3\bar{v}\nonumber\\
&+\bar{\alpha}_2\hbar^2\frac{\partial^2 \bar{v}}{\partial u^2}+2\hbar\bar{\alpha}_3\bar{v}+2\hbar\bar{\alpha}_3u\frac{\partial \bar{v}}{\partial u}+\pazocal{L}_{\mathrm{2LS}}\bar{v}.
\end{align}
\noindent
In the next step, we evaluate the operators $\mathcal{P}\hat{L}_2 \hat{L}_1^{-1}\hat{L}_2 \bar{v}$ and $\mathcal{P}\hat{L}_2 \hat{L}_1^{-1}(\dotsb)\bar{v}$. 
As a first step, we apply $\hat{L}_2$ to $\bar{v}$ to obtain
\begin{align}\label{a2}
\hat{L}_2 \bar{v}=-\biggl(\frac{i\hbar}{m\gamma}\frac{\partial }{\partial u}+\frac{i}{\hbar}m\omega^2u\biggl)P_1(v)\bar{\rho}(u).
\end{align}
\noindent
Noting that when $\hat{L}_1$ acts on $P_1(v)$, we obtain
\begin{equation}
    \hat{L}_1P_1(v)= - \gamma P_1(v).
\end{equation}
One can now show that the application of $\hat{L}_1^{-1}$ into Eq.~(\ref{a2}) yields
\begin{subequations}\label{xgadd}
\begin{align}
&\hat{L}_1^{-1}\hat{L}_2 \bar{v}=\frac{1}{\gamma}\biggl(\frac{i\hbar}{m\gamma}\frac{\partial }{\partial u}+\frac{i}{\hbar}m\omega^2u\biggl)P_1(v)\bar{\rho}(u),\\
&\hat{L}_1^{-1}\biggl(v\hat{m}_4+\hat{m}_2\frac{\partial}{\partial v}\biggl)\bar{v}=\frac{1}{\gamma}\biggl(\frac{\hat{m}_2}{\gamma} -\hat{m}_4\biggl)vP_0(v)\bar{\rho}(u).
\end{align}
\end{subequations}
\noindent
In Eq.~(\ref{xgadd}), we apply $\hat{L}_2$ again, which yields
\begin{subequations}\label{eqnss}
\begin{align}
\hat{L}_2P_1(v)&=\Bigl(\sqrt{2\gamma}P_2(v)+\sqrt{\gamma}P_0(v)\Bigl)\biggl(-\frac{i}{\hbar}m\omega^2u\biggl)\nonumber\\
&-\sqrt{\frac{2}{\gamma}}P_2(v)\biggl(\frac{i\hbar}{m}\frac{\partial }{\partial u}\biggl),\\
\mathcal{P}\hat{L}_2\hat{L}_1^{-1}\hat{L}_2 \bar{v}&=P_0(v)\biggl[\biggl(\frac{\omega^2}{\gamma}\biggl)u\frac{\partial}{\partial u}\bar{\rho}(u)+\biggl(\frac{m^2\omega^4}{\hbar^2}\biggl)u^2\bar{\rho}(u)\biggl],\label{eqnss_r}
\end{align}
\end{subequations}
\noindent
and 
\begin{align}\label{eqnssx}
\mathcal{P}\hat{L}_2\hat{L}_1^{-1}&\biggl(v\hat{m}_4+\hat{m}_2\frac{\partial}{\partial v}\biggl)\bar{v}\nonumber\\
&= P_0(v)\biggl(\frac{im\omega^2}{\gamma\hbar}u\biggl)\Bigl(\gamma\hat{m}_4-\hat{m}_2\Bigl)\bar{\rho}(u).
\end{align}
\noindent
Substituting Eqns.~(\ref{eqnss_r})-(\ref{eqnssx}) back into Eq.~(\ref{felimp}), and using $\bar{v}=P_0(v)\bar{\rho}(u)$, Eq.~(\ref{felimp}) becomes:
\begin{align}\label{finme}
 & \frac{\partial}{\partial t} \bar{\rho}(u) = -(\bar{\alpha}_1)^{-1}\biggl[\biggl(\frac{\omega^2}{\gamma}\biggl)u\frac{\partial}{\partial u}+\biggl(\frac{m^2\omega^4}{\hbar^2}\biggl)u^2\biggl]\bar{\rho}\nonumber\\
 &-(\bar{\alpha}_1)^{-1}\biggl(\frac{im\omega^2}{\gamma \hbar}u\biggl)\bigl[\gamma\hat{m}_4-\hat{m}_2\bigl]\bar{\rho}+\hat{m}_1\frac{\partial}{\partial u}\bar{\rho}+u\hat{m}_3\bar{\rho}\nonumber\\
 &+\bar{\alpha}_2\hbar^2\frac{\partial^2}{\partial u^2}\bar{\rho}+2\hbar\bar{\alpha}_3\bar{\rho}+2\bar{\alpha}_3\hbar u\frac{\partial}{\partial u}\bar{\rho}+\pazocal{L}_\mathrm{2LS}\bar{\rho}.
\end{align}
\noindent
We have eliminated the fast variable $v$ by assuming it relaxes quickly when $\bar{\alpha}_1$ is large. Based on our assumption that $\bar{\alpha}_1$ is significantly larger than all the system parameters, and neglecting terms that scale as an order of $||\hat{m}_i|| / \bar{\alpha}_1$,  we reduce Eq.~(\ref{finme}) to:
\begin{align}\label{finme1}
 & \frac{\partial}{\partial t} \bar{\rho}(u) \approx -\biggl(\frac{\omega^2}{\bar{\alpha}_1\gamma}\biggl)u\frac{\partial}{\partial u}\bar{\rho}-\biggl(\frac{m^2\omega^4}{\bar{\alpha}_1\hbar^2}\biggl)u^2\bar{\rho}+\hat{m}_1\frac{\partial}{\partial u}\bar{\rho}\nonumber\\
&+u\hat{m}_3\bar{\rho}+\bar{\alpha}_2\hbar^2\frac{\partial^2}{\partial u^2}\bar{\rho}+2\hbar\bar{\alpha}_3\bar{\rho}+2\bar{\alpha}_3\hbar u\frac{\partial}{\partial u}\bar{\rho}+\pazocal{L}_\mathrm{2LS}\bar{\rho}.
\end{align}
\noindent
After performing adiabatic elimination, we found that $v=0$, which in turn yields $u=x$, leading to the following hybrid quantum-classical master equation for the diagonal elements:
\begin{align}\label{oqbm_mee}
    \frac{\partial}{\partial t} \bar{\rho}(x) &\approx -\lambda_1x\frac{\partial}{\partial x }\bar{\rho}-\lambda_2 x^2 \bar{\rho}+\hat{m}_1 \frac{\partial}{\partial x }\bar{\rho}+x\hat{m}_3\bar{\rho}\nonumber\\
    &+\lambda_3\frac{\partial^2}{\partial x^2 }\bar{\rho}+\lambda_4\bar{\rho}+\lambda_4 x \frac{\partial}{\partial x }\bar{\rho} + \pazocal{L}_\mathrm{2LS}\bar{\rho},
    \end{align}
 \noindent
 where the parameters are
\begin{align}
    \lambda_1=\frac{\omega^2}{\bar{\alpha}_1\gamma}, \hspace{2mm} \lambda_2=\frac{m^2\omega^4}{\bar{\alpha}_1\hbar^2},\hspace{2mm} \lambda_3=\bar{\alpha}_2\hbar^2, \hspace{2mm} \lambda_4=2\bar{\alpha}_3\hbar.
\end{align}
 \noindent
The master equation~(\ref{oqbm_mee}) concludes the derivation of Eq.~(\ref{oqbm_me}). 

\vspace{2mm}

\section{Equations for the cumulants}\label{cumulants}

Continuing from Eq.~(\ref{cum_eq1}), we present the remaining cumulants for the case $\langle x^n \rangle_c=0$ for $n\geq 3$; these are summarized in Table~\ref{tab:appendix_eqns1} below:
\begin{table*}[htbp]
\centering
\caption{List of cumulants where $\langle x^n\rangle_c =0$ for $n\geq 3$.}
\label{tab:appendix_eqns1}
\renewcommand{\arraystretch}{1.5} 
\begin{tabular}{c|l}
\hline\hline
Eq. & Expression \\
\hline
(1) & 
$ \begin{aligned}
\frac{d }{d t}\langle z \rangle_c 
&= -2\bar{\lambda}_2 \langle z^2 \rangle_c\langle z\rangle_c 
   -\Delta_1\langle z \rangle_c 
   -4\bigl\{\bar{\delta}_2+\bar{a}_7\langle i^2 \rangle_c\bigr\}e^{\langle i^0\rangle_c-\langle z^0\rangle_c} -4\bigl\{\langle i \rangle_c-\langle z \rangle_c\bigr\}\bigl\{\Omega+\bar{a}_7\langle i \rangle_c \bigr\}e^{\langle i^0\rangle_c - \langle z^0 \rangle_c}\\
&\quad -\bar{\beta}\bigl\{\langle x \rangle_c-\langle z \rangle_c\bigr\} e^{1-\langle z^0\rangle_c}  +4\bigl\{\bar{\delta}_3+\bar{a}_8\langle r^2 \rangle_c\bigr\}e^{\langle r^0\rangle_c - \langle z^0 \rangle_c} 
   +4\bar{a}_8\bigl\{\langle r \rangle_c-\langle z \rangle_c\bigr\}\langle r\rangle_c e^{\langle r^0\rangle_c-\langle z^0\rangle_c}
\end{aligned}$ \\[4ex]
\hline
(2) & $\begin{aligned}
    \frac{d }{d t}\langle z^2 \rangle_c &=-4\bar{\lambda}_2 \langle z^2 \rangle_c^2-2\Delta_1\langle z^2 \rangle_c+2\bar{\lambda}_3-8\bigl\{\bar{\delta}_2+\bar{a}_7\langle i^2\rangle_c\bigl\}\bigl\{\langle i\rangle_c-\langle z\rangle_c\bigl\}e^{\langle i^0\rangle_c-\langle z^0\rangle_c}-4\bigl\{\Omega+\bar{a}_7\langle i \rangle_c\bigl\}\\
&\quad \times\bigl\{\bigl(\langle i\rangle_c-\langle z\rangle_c\bigl)^2+\langle i^2\rangle_c-\langle z^2\rangle_c\bigl\}e^{\langle i^0\rangle_c-\langle z^0\rangle_c}+8\bigl\{\bar{\delta}_3+\bar{a}_8\langle r^2\rangle_c\bigl\}\bigl\{\langle r\rangle_c-\langle z\rangle_c\bigl\}e^{\langle r^0\rangle_c-\langle z^0\rangle_c}\\
&\quad +4\bar{a}_8\langle r\rangle_c\bigl\{\bigl(\langle r\rangle_c-\langle z\rangle_c\bigl)^2+\langle r^2\rangle_c-\langle z^2\rangle_c\bigl\}e^{\langle r^0\rangle_c-\langle z^0\rangle_c}-\bar{\beta}\bigl\{\bigl(\langle x\rangle_c-\langle z\rangle_c\bigl)^2+\chi-\langle x\rangle_c^2-\langle z^2\rangle_c\bigl\}e^{1-\langle z^0\rangle_c}
\end{aligned}$\\[6ex]
\hline
(3) & 
$\begin{aligned}
    \frac{d }{d t}\langle r^0 \rangle_c =-\bar{\lambda}_2 \langle r \rangle_c^2-\bar{\lambda}_2 \langle r^2 \rangle_c-\Delta_1+\Delta_3-\bar{a}_8\langle z \rangle_ce^{\langle r^0 \rangle_c - \langle z^0 \rangle_c}
\end{aligned}$\\[2ex] 
\hline
(4) & $\begin{aligned}
    \frac{d }{d t}\langle r \rangle_c = -2\bar{\lambda}_2 \langle r \rangle_c\langle r^2 \rangle_c-\Delta_1\langle r \rangle_c-\frac{1}{2}\bar{a}_2e^{1-\langle r^0 \rangle_c}-\bar{\delta}_3e^{\langle z^0 \rangle_c-\langle r^0\rangle_c}+\bar{a}_8\bigl\{\langle r \rangle_c\langle z \rangle_c-\langle z \rangle_c^2-\langle z^2 \rangle_c\bigl\}e^{\langle z^0\rangle_c-\langle r^0\rangle_c}
\end{aligned}$\\[2ex]
\hline
(5) & $\begin{aligned}
    \frac{d }{d t}\langle r^2 \rangle_c & = -2\bar{\lambda}_2\langle r \rangle_c^2-2\Delta_1\langle r^2 \rangle_c+2\bar{\lambda}_3-\bigl\{\bar{a}_8\langle r \rangle_c^2\langle z \rangle_c-2\bar{\delta}_3\bigl(\langle r \rangle_c-\langle z \rangle_c\bigl)-\bar{a}_8\langle r^2 \rangle_c\langle z \rangle_c-2\bar{a}_8\langle r \rangle_c\langle z \rangle_c^2+\bar{a}_8\langle z \rangle_c^3\nonumber\\
&\quad-2\bar{a}_8\langle r \rangle_c\langle z^2 \rangle_c+3\bar{a}_8\langle z \rangle_c\langle z^2 \rangle_c\bigl\}e^{\langle z^0\rangle_c-\langle r^0\rangle_c}+\bar{a}_2\bigl\{\langle r \rangle_c-\langle x \rangle_c\bigl\}e^{1-\langle r^0 \rangle_c}
\end{aligned}$\\[4ex]
\hline
(6) & $\begin{aligned} 
\frac{d }{d t}\langle i^0 \rangle_c =-\bar{\lambda}_2 \langle i\rangle_c^2-\bar{\lambda}_2 \langle i^2 \rangle_c+\Delta_4-\Delta_1+\bigl\{\Omega+\bar{a}_2\langle z\rangle_c\bigl\}e^{\langle r^0\rangle_c-\langle i^0\rangle_c}
\end{aligned}
$\\[2ex]
\hline
(7) & $\begin{aligned} 
\frac{d }{d t}\langle i \rangle_c &=-2\bar{\lambda}_2 \langle i\rangle_c\langle i^2\rangle_c-\Delta_1\langle i\rangle_c+\frac{1}{2}\bar{\delta}_1e^{1-\langle i^0\rangle_c}+\bigl\{\Omega+\bar{a}_2\langle z\rangle_c\bigl\}\bigl\{\langle i\rangle_c-\langle z \rangle_c\bigl\}e^{\langle z^0\rangle_c-\langle i^0 \rangle_c}+\bigl\{\bar{\delta}_2+\bar{a}_2\langle z^2\rangle_c\bigl\}e^{\langle z^0\rangle_c-\langle i^0\rangle_c}
\end{aligned}
$\\[2ex]
\hline
(8) & $\begin{aligned} 
\frac{d }{d t}\langle i^2 \rangle_c &=-2\bar{\lambda}_2\langle i^2\rangle_c^2-2\Delta_1\langle i^2\rangle_c+2\bar{\lambda}_3+\bigl\{\Omega+\bar{a}_2\langle z\rangle_c\bigl\}\bigl\{\bigl(\langle i\rangle_c-\langle z \rangle_c\bigl)^2+\langle z^2\rangle_c-\langle i^2\rangle_c\bigl\}e^{\langle z^0\rangle_c-\langle i^0\rangle_c}-2\bigl\{\bar{\delta}_2+\bar{a}_2\langle z^2\rangle_c\bigl\}\nonumber\\
&\quad\times\bigl\{\langle i \rangle_c - \langle z\rangle_c\bigl\}e^{\langle z^0\rangle_c - \langle i^0\rangle_c}
\end{aligned}
$\\[4ex]
\hline
\end{tabular}
\end{table*}
The remaining cumulants, continuing from Eqn~(\ref{4th_zerocumulants}) for the case $ \langle x^n\rangle_c=0$ for $n\geq4$ are summarized in Table~\ref{tab:appendix_eqns2}.
\begin{table*}[htbp]
\centering
\caption{List of cumulants where $\langle x^n\rangle_c =0$ for $n\geq 4$.}
\label{tab:appendix_eqns2}
\renewcommand{\arraystretch}{1.5} 
\begin{tabular}{c|l}
\hline\hline
Eq. & Expression \\
\hline 
(1) & 
$ \begin{aligned}
\frac{d }{d t}\langle z^0 \rangle_c =-\bar{\lambda}_2 \langle z^2 \rangle_c-\bar{\lambda}_2 \langle z \rangle_c^2-\Delta_1+\Delta_4-4\bigl\{\Omega+\bar{a}_7\langle i \rangle_c\bigl\}e^{\langle i^0\rangle_c-\langle z^0\rangle_c}-\bar{\beta}e^{1-\langle z^0\rangle_c}+4\bar{a}_8\langle r \rangle_c e^{\langle r^0\rangle_c-\langle z^0 \rangle_c}
\end{aligned}$ \\[2ex]
\hline
(2) & 
$ \begin{aligned}
\frac{d }{d t}\langle z \rangle_c &=-\bar{\lambda}_2\langle z^3 \rangle_c-2\bar{\lambda}_2\langle z \rangle_c\langle z^2 \rangle_c-\Delta_1\langle z \rangle_c-4\bigl\{\bar{\delta}_2+\Omega\langle i \rangle_c+\bar{a}_7\langle i \rangle_c^2+\bar{a}_7\langle i^2 \rangle_c-\Omega\langle z \rangle_c-\bar{a}_7\langle i \rangle_c\langle z \rangle_c\bigl\}e^{\langle i^0 \rangle_c-\langle z^0 \rangle_c}\\
&\quad+4\bigl\{\bar{\delta}_3+\bar{a}_8\langle r \rangle_c^2+\bar{a}_8\langle r^2 \rangle_c -\bar{a}_8\langle r \rangle_c\langle z \rangle_c\bigl\}e^{\langle r^0 \rangle_c-\langle z^0 \rangle_c}+\bar{\beta}\bigl\{\langle z \rangle_c-\langle x \rangle_c\bigl\}e^{1-\langle z^0 \rangle_c}
\end{aligned}$ \\[4ex]
\hline
(3) & 
$ \begin{aligned}
\frac{d }{d t}\langle z^2 \rangle_c &= -2\bar{\lambda}_2\langle z^2 \rangle_c^2-2\bar{\lambda}_2\langle z \rangle_c\langle z^3 \rangle_c-2\Delta_1\langle z^2 \rangle_c+2\bar{\lambda}_3-4\bar{a}_7\langle i^3 \rangle_ce^{\langle i^0 \rangle_c-\langle z^0 \rangle_c}-8\bigl\{\bar{\delta}_2+\bar{a}_7\langle i^2\rangle_c\bigl\}\bigl\{\langle i\rangle_c-\langle z\rangle_c\bigl\}\\
    &\quad\times e^{\langle i^0\rangle_c-\langle z^0\rangle_c}-4\bigl\{\Omega+\bar{a}_7\langle i\rangle_c\bigl\}\bigl\{\langle i\rangle_c^2+\langle i^2\rangle_c-2\langle i\rangle_c\langle z\rangle_c+\langle z\rangle_c^2-\langle z^2\rangle_c\bigl\}e^{\langle i^0\rangle_c-\langle z^0\rangle_c}+4\bigl\{2\bar{\delta}_3\langle r\rangle_c+\bar{a}_8\langle r\rangle_c^3\\
    &\quad+3\bar{a}_8\langle r\rangle_c\langle r^2\rangle_c+\bar{a}_8\langle r^3\rangle_c-2\bar{\delta}_3\langle z\rangle_c-2\bar{a}_8\langle r\rangle_c^2\langle z\rangle_c-2\bar{a}_8\langle r^2\rangle_c\langle z\rangle_c+\bar{a}_8\langle r\rangle_c\langle z\rangle_c^2-\bar{a}_8\langle r\rangle_c\langle z^2\rangle_c\bigl\}e^{\langle r^0\rangle_c-\langle z^0\rangle_c}
\end{aligned}$ \\[6ex]
\hline
(4) & 
$ \begin{aligned}
\frac{d }{d t}\langle z^3 \rangle_c &=-3\Delta_1\langle z^3\rangle_c - 6\bar{\lambda}_2\langle z^2\rangle_c\langle z^3\rangle_c-12\bar{a}_7\langle i^3\rangle_c\bigl\{\langle i\rangle_c-\langle z\rangle_c\bigl\}e^{\langle i^0\rangle_c-\langle z^0\rangle_c}+12\bar{a}_8\langle r^3\rangle_c\bigl\{\langle r\rangle_c-\langle z\rangle_c\bigl\}e^{\langle r^0\rangle_c-\langle z^0\rangle_c}\\
&\quad-12\bigl\{\bar{\delta}_2+\bar{a}_7\langle i^2\rangle_c\bigl\}\bigl\{\langle i\rangle_c^2+\langle i^2\rangle_c-2\langle i\rangle_c\langle z\rangle_c+\langle z\rangle_c^2-\langle z^2\rangle_c\bigl\}e^{\langle i^0\rangle_c-\langle z^0\rangle_c}+12\bigl\{\bar{\delta}_3+\bar{a}_8\langle r^2\rangle_c\bigl\}\bigl\{\langle r\rangle_c^2+\langle r^2\rangle_c\\
    &\quad-2\langle r\rangle_c\langle z\rangle_c+\langle z\rangle_c^2-\langle z^2\rangle_c\bigl\}e^{\langle r^0\rangle_c-\langle z^0\rangle_c}-4\bigl\{\Omega+\bar{a}_7\langle i\rangle_c\bigl\}\Bigl\{\langle i\rangle_c^3+\langle i^3\rangle_c-3\langle i\rangle_c^2\langle z\rangle_c-3\langle i^2\rangle_c\langle z\rangle_c-\langle z\rangle_c^3\\
    &\quad+3\langle i\rangle_c\bigl\{\langle i^2\rangle_c +\langle z\rangle_c^2-\langle z^2\rangle_c\bigl\}+3\langle z\rangle_c\langle z^2\rangle_c-\langle z^3\rangle_c\Bigl\}e^{\langle i^0\rangle_c-\langle z^0\rangle_c}+4\bar{a}_8\langle r \rangle_c\Bigl\{\langle r\rangle_c^3+\langle r^3\rangle_c-3\langle r\rangle_c^2\langle z\rangle_c-\langle z\rangle_c^3\\
    &\quad+3\langle r\rangle_c\bigl\{\langle r^2\rangle_c+\langle z\rangle_c^2-\langle z^2\rangle_c\bigl\}+3\langle z\rangle_c\langle z^2\rangle_c-\langle z^3\rangle_c\Bigl\}e^{\langle r^0\rangle_c-\langle z^0\rangle_c}+\bar{\beta}\Bigl\{ \langle z\rangle_c^3+\langle z^3\rangle_c-3\langle z\rangle_c^2\langle x\rangle_c+3\langle z^2\rangle_c\langle x\rangle_c\\
    &\quad+2\langle x\rangle_c^3-3\chi\langle x\rangle_c+3\langle z\rangle_c\bigl\{\chi-\langle z^2\rangle_c\bigl\}\Bigl\}e^{1-\langle z^0\rangle_c}
\end{aligned}$ \\[8ex]
\hline 
(5) & 
$ \begin{aligned}
\frac{d }{d t}\langle r^0 \rangle_c &=-\bar{\lambda}_2\langle r \rangle_c^2-\bar{\lambda}_2\langle r^2 \rangle_c-\Delta_1+\Delta_3-\bar{a}_8\langle z \rangle_ce^{\langle z^0 \rangle_c-\langle r^0 \rangle_c}
\end{aligned}$ \\[2ex]
\hline 
(6) & 
$ \begin{aligned}
\frac{d }{d t}\langle r \rangle_c &=-\bar{\lambda}_2\langle r^3 \rangle_c-2\bar{\lambda}_2\langle r \rangle_c\langle r^2 \rangle_c-\Delta_1\langle r \rangle_c-\frac{1}{2}\bar{a}_2e^{1-\langle r^0 \rangle_c}-\bigl\{\bar{\delta}_3-\bar{a}_8\langle r \rangle_c\langle z \rangle_c+\bar{a}_8\langle z \rangle_c^2+\bar{a}_8\langle z^2 \rangle_c\bigl\}e^{\langle z^0 \rangle_c-\langle r^0 \rangle_c}
\end{aligned}$ \\[2ex]
\hline 
(7) & 
$ \begin{aligned}
\frac{d }{d t}\langle r^2 \rangle_c &=-2\bar{\lambda}_2\langle r^2 \rangle_c^2-2\bar{\lambda}_2\langle r \rangle_c\langle r^3 \rangle_c-2\Delta_1\langle r^2 \rangle_c+2\bar{\lambda}_3-\bigl\{2\bar{\delta}_3\langle z \rangle_c-2\bar{\delta}_3\langle r \rangle_c+\bar{a}_8\langle r \rangle_c^2\langle z \rangle_c-\bar{a}_8\langle r^2 \rangle_c\langle z \rangle_c\\
&\quad-2\bar{a}_8\langle r \rangle_c\langle z \rangle_c^2+\bar{a}_8\langle z \rangle_c^3-2\bar{a}_8\langle r \rangle_c\langle z^2 \rangle_c+3\bar{a}_8\langle z \rangle_c\langle z^2 \rangle_c+\bar{a}_8\langle z^3 \rangle_c\bigl\}e^{\langle z^0 \rangle_c-\langle r^0 \rangle_c}+\bar{a}_2\bigl\{\langle r \rangle_c-\langle x \rangle_c\bigl\}e^{1-\langle r^0 \rangle_c}
\end{aligned}$ \\[4ex]
\hline
(8) & 
$ \begin{aligned}
\frac{d }{d t}\langle r^3 \rangle_c &=-3\Delta_1\langle r^3 \rangle_c-6\bar{\lambda}_2\langle r^2 \rangle_c\langle r^3 \rangle_c-3\bigl\{\bar{\delta}_3+\bar{a}_8\langle z^2 \rangle_c\bigl\}\bigl\{\langle r \rangle_c^2-\langle r^2 \rangle_c-2\langle r \rangle_c\langle z \rangle_c+\langle z \rangle_c^2+\langle z^2 \rangle_c\bigl\}e^{\langle z^0 \rangle_c-\langle r^0 \rangle_c}\\
&\quad+\bar{a}_8\langle z \rangle_c\bigl\{\langle r \rangle_c^3-3\langle r \rangle_c\langle r^2 \rangle_c+\langle r^3 \rangle_c-3\langle r \rangle_c^2\langle z \rangle_c+3\langle r^2 \rangle_c\langle z \rangle_c+3\langle r \rangle_c\langle z \rangle_c^2-\langle z \rangle_c^3+3\langle r \rangle_c\langle z^2 \rangle_c-3\langle z \rangle_c\langle z^2 \rangle_c\\
&\quad-\langle z^3 \rangle_c\bigl\}e^{\langle z^0 \rangle_c-\langle r^0 \rangle_c}+3\bar{a}_8\bigl\{\langle r \rangle_c-\langle z \rangle_c\bigl\}\langle z^3 \rangle_ce^{\langle z^0 \rangle_c-\langle r^0 \rangle_c}-\frac{3}{2}\bar{a}_2\bigl\{\langle r \rangle_c^2-\langle r^2 \rangle_c-2\langle r \rangle_c\langle x \rangle_c+\chi\bigl\}e^{1-\langle r^0 \rangle_c}
\end{aligned}$ \\[6ex]
\hline
(9) & 
$ \begin{aligned}
\frac{d }{d t}\langle i^0 \rangle_c =-\bar{\lambda}_2\langle i \rangle_c^2-\bar{\lambda}_2\langle i^2 \rangle_c+\Delta_4-\Delta_1+\bigl\{\Omega+\bar{a}_2\langle z \rangle_c\bigl\}e^{\langle z^0 \rangle_c-\langle i^0 \rangle_c}
\end{aligned}$ \\[2ex]
\hline 
(10) & 
$ \begin{aligned}
\frac{d }{d t}\langle i \rangle_c &=-\bar{\lambda}_2\langle i^3 \rangle_c-2\bar{\lambda}_2\langle i \rangle_c\langle i^2 \rangle_c-\Delta_1\langle i \rangle_c+\frac{1}{2}\bar{\delta}_1e^{1-\langle i^0 \rangle_c}-\bigl\{\langle i \rangle_c-\langle z \rangle_c\bigl\}\bigl\{\Omega+\bar{a}_2\langle z \rangle_c\bigl\}e^{\langle z^0 \rangle_c-\langle i^0 \rangle_c}\\
&\quad+\bigl\{\bar{\delta}_2+\bar{a}_2\langle z^2 \rangle_c\bigl\}e^{\langle i^0 \rangle_c-\langle z^0 \rangle_c}
\end{aligned}$ \\[2ex]
\hline 
(11) & 
$ \begin{aligned}
\frac{d }{d t}\langle i^2 \rangle_c &=-2\bar{\lambda}_2\langle i^2 \rangle_c^2-2\bar{\lambda}_2\langle i \rangle_c\langle i^3 \rangle_c-2\Delta_1\langle i^2 \rangle_c+2\bar{\lambda}_3+\bigl\{\Omega+\bar{a}_2\langle z \rangle_c\bigl\}\bigl\{\langle i \rangle_c^2-\langle i^2 \rangle_c-2\langle i \rangle_c\langle z \rangle_c+\langle z \rangle_c^2+\langle z^2 \rangle_c\bigl\}\\
    &\quad\times e^{\langle z^0 \rangle_c-\langle i^0 \rangle_c}-2\bigl\{\langle i \rangle_c-\langle z \rangle_c\bigl\}\bigl\{\bar{\delta}_2+\bar{a}_2\langle z^2 \rangle_c\bigl\}e^{\langle z^0 \rangle_c-\langle i^0 \rangle_c}+\bar{a}_2\langle z^3 \rangle_ce^{\langle z^0 \rangle_c-\langle i^0 \rangle_c}-\bar{\delta}_1\bigl\{\langle i \rangle_c-\langle x \rangle_c\bigl\}e^{1-\langle i^0 \rangle_c}
\end{aligned}$ \\[4ex]
\hline 
(12) & 
$ \begin{aligned}
\frac{d }{d t}\langle i^3\rangle_c &=-3\Delta_1\langle i^3 \rangle_c-6\bar{\lambda}_2\langle i^2 \rangle_c\langle i^3 \rangle_c+3\bigl\{\bar{\delta}_2+\bar{a}_2\langle z^2 \rangle_c\bigl\}\bigl\{\langle i \rangle_c^2-\langle i^2 \rangle_c-2\langle i \rangle_c\langle z \rangle_c+\langle z \rangle_c^2+\langle z^2 \rangle_c\bigl\}e^{\langle z^0 \rangle_c-\langle i^0 \rangle_c}\\
&\quad-3\bar{a}_2\langle z^3 \rangle_c\bigl\{\langle i \rangle_c-\langle z \rangle_c\bigl\}e^{\langle z^0 \rangle_c-\langle i^0 \rangle_c}+\bigl\{\Omega+\bar{a}_2\langle z \rangle_c\bigl\}\bigl\{3\langle i \rangle_c^2\langle z \rangle_c-\langle i \rangle_c^3-\langle i^3 \rangle_c-3\langle i^2 \rangle_c\langle z \rangle_c+\langle z \rangle_c^3+3\langle i \rangle_c\langle i^2 \rangle_c\\
&\quad-3\langle i \rangle_c\langle z \rangle_c^2-3\langle i \rangle_c\langle z^2 \rangle_c+3\langle z \rangle_c\langle z^2 \rangle_c+\langle z^3 \rangle_c\bigl\}e^{\langle z^0 \rangle_c-\langle i^0 \rangle_c}+\frac{3}{2}\bar{\delta}_1\bigl\{\langle i \rangle_c^2-\langle i^2 \rangle_c-2\langle i \rangle_c\langle x \rangle_c+\chi\bigl\}e^{1-\langle i^0 \rangle_c}
\end{aligned}$ \\[6ex]
\hline 
\end{tabular}
\end{table*}
\nocite{*}
\bibliography{oqBm2}

@PREAMBLE{
 "\providecommand{\noopsort}[1]{}" 
 # "\providecommand{\singleletter}[1]{#1}%" 
}

@article{kramers1940brownian,
title = {Brownian motion in a field of force and the diffusion model of chemical reactions},
journal = {Physica},
volume = {7},
number = {4},
pages = {284-304},
year = {1940},
issn = {0031-8914},
doi = {https://doi.org/10.1016/S0031-8914(40)90098-2},
url = {https://www.sciencedirect.com/science/article/pii/S0031891440900982},
author = {H.A. Kramers},
}

@book{breuer2002theory,
  title={{The Theory of Open Quantum Systems}},
  author={Breuer, Heinz-Peter and Petruccione, Francesco},
  year={2002},
  publisher={Oxford University Press, Oxford}
}

@book{schlosshauer2007decoherence,
  title={{Decoherence and the Quantum-To-Classical Transition, The Frontiers Collection}},
  author={Schlosshauer, Maximilian A},
  year={2007},
  publisher={Springer}
}

@book{nielsen2010quantum,
  title={Quantum computation and quantum information},
  author={Nielsen, Michael A and Chuang, Isaac L},
  year={2010},
  publisher={Cambridge university press}
}

@article{smoluchowski1916brownsche,
  title={{{\"U}ber Brownsche Molekularbewegung unter Einwirkung {\"a}u{\ss}erer Kr{\"a}fte und deren Zusammenhang mit der verallgemeinerten Diffusionsgleichung}},
  author={Smoluchowski, Marian V},
  journal={Annalen der Physik},
  volume={353},
  number={24},
  pages={1103--1112},
  year={1916},
  publisher={WILEY-VCH Verlag Leipzig}
}

@article{ATTAL20121545,
title = {Open quantum walks on graphs},
journal = {Phys. Lett. A.},
volume = {376},
number = {18},
pages = {1545-1548},
year = {2012},
issn = {0375-9601},
doi = {https://doi.org/10.1016/j.physleta.2012.03.040},
url = {https://www.sciencedirect.com/science/article/pii/S0375960112003453},
author = {S. Attal and F. Petruccione and I. Sinayskiy}
}

@article{attal2012open,
  title={Open quantum random walks},
  author={Attal, Stephane and Petruccione, Francesco and Sabot, Christophe and Sinayskiy, Ilya},
  journal={J. Stat. Phys.},
  volume={147},
  number={4},
  pages={832--852},
  year={2012},
doi = {https://doi.org/10.1007/s10955-012-0491-0},
  publisher={Springer}
}

@article{Sinayskiy_2012,
doi = {10.1088/0031-8949/2012/T151/014077},
url = {https://dx.doi.org/10.1088/0031-8949/2012/T151/014077},
year = {2012},
month = {nov},
publisher = {IOP Publishing},
volume = {2012},
number = {T151},
pages = {014077},
author = {Ilya Sinayskiy and Francesco Petruccione},
title = {Properties of open quantum walks on $\mathbb{Z}$},
journal = {Phys. Scr}
}

@book{kraus1983states,
  title={{States, Effects, and Operations Fundamental Notions of Quantum Theory: Lectures in Mathematical Physics at the University of Texas at Austin}},
  author={Kraus, Karl and B{\"o}hm, Arno and Dollard, John D and Wootters, WH},
  year={1983},
  publisher={Springer}
}

@article{sinayskiy2012efficiency,
  title={Efficiency of open quantum walk implementation of dissipative quantum computing algorithms},
  author={Sinayskiy, Ilya and Petruccione, Francesco},
  journal={Quant. Inf. Proc.},
  volume={11},
  pages={1301--1309},
  year={2012},
  publisher={Springer},
doi = {https://doi.org/10.1007/s11128-012-0426-3}
}

@article{lindblad1976generators,
  title={On the generators of quantum dynamical semigroups},
  author={Lindblad, Goran},
  journal={Commun. Math. Phys.},
  volume={48},
  pages={119--130},
  year={1976},
doi={https://doi.org/10.1007/BF01608499},
  publisher={Springer}
}

@article{gorini1976completely,
  title={{Completely positive dynamical semigroups of $N$-level systems}},
  author={Gorini, Vittorio and Kossakowski, Andrzej and Sudarshan, Ennackal Chandy George},
  journal={J. Math. Phys.},
  volume={17},
  number={5},
  pages={821--825},
  year={1976},
  publisher={American Institute of Physics},
doi={https://doi.org/10.1063/1.522979}
}

@inproceedings{attal2015central,
  title={Central limit theorems for open quantum random walks and quantum measurement records},
  author={Attal, St{\'e}phane and Guillotin-Plantard, Nadine and Sabot, Christophe},
  booktitle={	Ann. Henri Poincaré},
  volume={16},
  pages={15--43},
  year={2015},
  organization={Springer},
doi={https://doi.org/10.1007/s00023-014-0319-3}
}

@article{konno2013limit,
  title={Limit theorems for open quantum random walks},
  author={Konno, Norio and Yoo, Hyun Jae},
  journal={J. Stat. Phys.},
  volume={150},
  pages={299--319},
  year={2013},
  publisher={Springer},
doi={https://doi.org/10.1007/s10955-012-0668-6}
}

@article{aharonov1993quantum,
  title = {Quantum random walks},
  author = {Aharonov, Y. and Davidovich, L. and Zagury, N.},
  journal = {Phys. Rev. A.},
  volume = {48},
  issue = {2},
  pages = {1687--1690},
  numpages = {0},
  year = {1993},
  month = {Aug},
  publisher = {American Physical Society},
  doi = {10.1103/PhysRevA.48.1687},
  url = {https://link.aps.org/doi/10.1103/PhysRevA.48.1687}
}

@article{kempe2003quantum,
author = {J Kempe},
title = {{Quantum random walks: An introductory overview}},
journal = {Contemp. Phys.},
volume = {44},
number = {4},
pages = {307-327},
year = {2003},
publisher = {Taylor & Francis},
doi = {https://doi.org/10.1080/00107151031000110776}

}

@article{pellegrini2014continuous,
  title={{Continuous time open quantum random walks and non-Markovian Lindblad master equations}},
  author={Pellegrini, Cl{\'e}ment},
  journal={J. Stat. Phys.},
  volume={154},
  number={3},
  pages={838--865},
  year={2014},
  publisher={Springer},
doi={https://doi.org/10.1007/s10955-013-0910-x}
}

@article{caldeira1983path,
title = {{Path integral approach to quantum Brownian motion}},
journal = {Phys. A.},
volume = {121},
number = {3},
pages = {587-616},
year = {1983},
issn = {0378-4371},
doi = {https://doi.org/10.1016/0378-4371(83)90013-4},
url = {https://www.sciencedirect.com/science/article/pii/0378437183900134},
author = {A.O. Caldeira and A.J. Leggett},
}

@article{CALDEIRA1983374,
title = {Quantum tunnelling in a dissipative system},
journal = {Ann. Phys. (NY).},
volume = {149},
number = {2},
pages = {374-456},
year = {1983},
issn = {0003-4916},
doi = {https://doi.org/10.1016/0003-4916(83)90202-6},
url = {https://www.sciencedirect.com/science/article/pii/0003491683902026},
author = {A.O Caldeira and A.J Leggett}
}

@article{van1985elimination,
title = {Elimination of fast variables},
journal = {Phys. Rep.},
volume = {124},
number = {2},
pages = {69-160},
year = {1985},
issn = {0370-1573},
doi = {https://doi.org/10.1016/0370-1573(85)90002-X},
url = {https://www.sciencedirect.com/science/article/pii/037015738590002X},
author = {N.G. {Van Kampen}},

}

@book{gardiner1985handbook,
  title={{Handbook of Stochastic Methods}},
  author={Gardiner, Crispin W and others},
  volume={3},
  year={1985},
  publisher={springer Berlin}
}

@article{bauer2013open,
  title = {{Open quantum random walks: Bistability on pure states and ballistically induced diffusion}},
  author = {Bauer, Michel and Bernard, Denis and Tilloy, Antoine},
  journal = {Phys. Rev. A.},
  volume = {88},
  issue = {6},
  pages = {062340},
  numpages = {6},
  year = {2013},
  month = {Dec},
  publisher = {American Physical Society},
  doi = {10.1103/PhysRevA.88.062340},
  url = {https://link.aps.org/doi/10.1103/PhysRevA.88.062340}
}

@article{bauer2014open,
doi = {10.1088/1742-5468/2014/09/P09001},
url = {https://dx.doi.org/10.1088/1742-5468/2014/09/P09001},
year = {2014},
month = {sep},
publisher = {IOP Publishing and SISSA},
volume = {2014},
number = {9},
pages = {P09001},
author = {Michel Bauer and Denis Bernard and Antoine Tilloy},
title = {{The open quantum Brownian motions}},
journal = {J. Stat. Mech.},

}

@article{sinayskiy2015microscopicbrown,
  title={{Microscopic derivation of open quantum Brownian motion: a particular example}},
  author={Sinayskiy, Ilya and Petruccione, Francesco},
  journal={Phys. Scr.},
  volume={2015},
  number={T165},
  pages={014017},
  year={2015},
doi={http://dx.doi.org/10.1088/0031-8949/2015/T165/014017},
  publisher={IOP Publishing}
}

@article{sinayskiy2017steady,
  title={{Steady-State control of open Quantum Brownian Motion}},
  author={Sinayskiy, Ilya and Petruccione, Francesco},
  journal={Fortschr. Phys.},
  volume={65},
  number={6-8},
  pages={1600063},
  year={2017},
  publisher={Wiley Online Library},
doi={https://doi.org/10.1002/prop.201600063}
}

@book{gardiner2004quantum,
  title={{Quantum Noise: A Handbook of Markovian and Non-Markovian Quantum Stochastic Methods with Applications to Quantum Optics, Springer Series in Synergetics}},
  author={Gardiner, Crispin and Zoller, Peter},
  year={2004},
  publisher={Springer, Berlin }
}

@article{zungu2025adiabatic,
  title={{Adiabatic elimination and Wigner function approach in microscopic derivation of Open Quantum Brownian Motion}},
  author={Zungu, Ayanda and Sinayskiy, Ilya and Petruccione, Francesco},
  journal={arXiv:2503.10379},
  year={2025},
url ={https://doi.org/10.48550/arXiv.2503.10379}
}

@book{carmichael1999statistical,
  title={{Statistical Methods in Quantum Optics 1: Master Equations and Fokker-Planck Equations}},
  author={Carmichael, Howard},
  volume={1},
  year={2002},
  publisher={Berlin: Springer}
}

@book{scully1999quantum,
  title={Quantum optics},
  author={Scully, Marlan O and Zubairy, M Suhail},
  year={1997},
  publisher={Cambridge: Cambridge University Press}
}

@article{L.Diósi_1993,
doi = {10.1209/0295-5075/22/1/001},
url = {https://dx.doi.org/10.1209/0295-5075/22/1/001},
year = {1993},
month = {apr},
publisher = {},
volume = {22},
number = {1},
pages = {1},
author = {L. Diósi},
title = {{On High-Temperature Markovian Equation for Quantum Brownian Motion}},
journal = {EPL.},
}

@article{diosi1993calderia,
  title={{Calderia-Leggett master equation and medium temperatures}},
  author={Di{\'o}si, Lajos},
  journal={Phys. A: Stat. Mech. Appl.},
  volume={199},
  number={3-4},
  pages={517--526},
  year={1993},
  publisher={Elsevier},
doi={https://doi.org/10.1016/0378-4371(93)90065-C}
}

@article{homa2019positivity,
  title={{Positivity violations of the density operator in the Caldeira-Leggett master equation}},
  author={Homa, G{\'a}bor and Bern{\'a}d, J{\'o}zsef Zsolt and Lisztes, L{\'a}szl{\'o}},
  journal={Eur. Phys. J. D.},
  volume={73},
  pages={1--13},
  year={2019},
  publisher={Springer},
doi={https://doi.org/10.1140/epjd/e2019-90604-4}
}

@article{vacchini2000completely,
  title={Completely positive quantum dissipation},
  author={Vacchini, Bassano},
  journal={Phys. Rev. Lett.},
  volume={84},
  number={7},
  pages={1374},
  year={2000},
  publisher={APS},
doi={https://doi.org/10.1103/PhysRevLett.84.1374}
}

@book{kardar2007statistical,
  title={Statistical physics of particles},
  author={Kardar, Mehran},
  year={2007},
  publisher={Cambridge University Press}
}

@article{thimmel1999rotating,
  title={Rotating wave approximation: systematic expansion and application to coupled spin pairs},
  author={Thimmel, Bernhard and Nalbach, Peter and Terzidis, Orestis},
  journal={Eur. Phys. J. B.},
  volume={9},
  pages={207--214},
  year={1999},
  publisher={Springer},
doi={https://doi.org/10.1007/s100510050758}
}

@article{raju2016quantum,
  title={{Quantum dissipative effects on
non-equilibrium transport through
a single-molecular transistor: The Anderson-Holstein-Caldeira-Leggett model}},
  author={Raju, Ch Narasimha and Chatterjee, Ashok},
  journal={Sci. Rep.},
  volume={6},
  number={1},
  pages={18511},
  year={2016},
  publisher={Nature Publishing Group UK London},
doi={https://doi.org/10.1038/srep18511}
}

@article{marquardt2004relaxation,
  title={{Relaxation and dephasing in a many-fermion generalization of the Caldeira-Leggett model}},
  author={Marquardt, Florian and Golubev, DS},
  journal={Phys. Rev. Lett.},
  volume={93},
  number={13},
  pages={130404},
  year={2004},
  publisher={APS},
doi={https://doi.org/10.1103/PhysRevLett.93.130404}
}

@article{yabu1998dissipative,
  title={{Dissipative field theory with the Caldeira-Leggett method and its application to disoriented chiral condensation}},
  author={Yabu, Hiroyuki and Nozawa, Kenji and Suzuki, Toru},
  journal={Phys. Rev. D.},
  volume={57},
  number={3},
  pages={1687},
  year={1998},
  publisher={APS},
doi={ https://doi.org/10.1103/PhysRevD.57.1687}
}

@article{bai2017classical,
  title={Classical-to-quantum transition behavior between two oscillators separated in space under the action of optomechanical interaction},
  author={Bai, Cheng-Hua and Wang, Dong-Yang and Wang, Hong-Fu and Zhu, Ai-Dong and Zhang, Shou},
  journal={Sci. Rep.},
  volume={7},
  number={1},
  pages={2545},
  year={2017},
  publisher={Nature Publishing Group UK London},
doi={https://doi.org/10.1038/s41598-017-02779-w}
}

@article{kovacs2017quantum,
  title={{Quantum--classical transition in the Caldeira--Leggett model}},
  author={Kov{\'a}cs, J{\'o}zsef and Fazekas, B and Nagy, S{\'a}ndor and Sailer, Korn{\'e}l},
  journal={Ann. Phys. (NY).},
  volume={376},
  pages={372--381},
  year={2017},
  publisher={Elsevier},
doi={https://doi.org/10.1016/j.aop.2016.12.010}
}

@article{layton2024healthier,
  title={A healthier semi-classical dynamics},
  author={Layton, Isaac and Oppenheim, Jonathan and Weller-Davies, Zachary},
  journal={Quantum},
  volume={8},
  pages={1565},
  year={2024},
  publisher={Verein zur F{\"o}rderung des Open Access Publizierens in den Quantenwissenschaften},
doi={https://doi.org/10.22331/q-2024-12-16-1565}
}

@article{layton2024classical,
  title={The classical-quantum limit},
  author={Layton, Isaac and Oppenheim, Jonathan},
  journal={PRX Quantum},
  volume={5},
  number={2},
  pages={020331},
  year={2024},
  publisher={APS},
doi={https://doi.org/10.1103/PRXQuantum.5.020331}
}

@article{oppenheim2023postquantum,
  title={A postquantum theory of classical gravity?},
  author={Oppenheim, Jonathan},
  journal={Phys. Rev. X.},
  volume={13},
  number={4},
  pages={041040},
  year={2023},
  publisher={APS},
doi={https://doi.org/10.1103/PhysRevX.13.041040}
}

@article{oppenheim2022two,
  title={The two classes of hybrid classical-quantum dynamics},
  author={Oppenheim, Jonathan and Sparaciari, Carlo and {\v{S}}oda, Barbara and Weller-Davies, Zachary},
  journal={arXiv:2203.01332},
  year={2022},
url={https://doi.org/10.48550/arXiv.2203.01332}
}

@article{halliwell1998effective,
  title={{Effective theories of coupled classical and quantum variables from decoherent histories: A New approach to the back reaction problem}},
  author={Halliwell, JJ},
  journal={Phys. Rev. D.},
  volume={57},
  number={4},
  pages={2337},
  year={1998},
  publisher={APS},
doi={https://doi.org/10.1103/PhysRevD.57.2337}
}

@article{PhysRevA.107.062206,
  title= {{Hybrid completely positive Markovian quantum-classical dynamics}},
  author= {Di\'osi, Lajos},
  journal= {Phys. Rev. A.},
  volume= {107},
  issue= {6},
  pages= {062206},
  numpages= {10},
  year= {2023},
  month= {Jun},
  publisher= {American Physical Society},
  doi= {10.1103/PhysRevA.107.062206},
  url= {https://link.aps.org/doi/10.1103/PhysRevA.107.062206}
}

@article{tilloy2024general,
  title={General quantum-classical dynamics as measurement based feedback},
  author={Tilloy, Antoine},
  journal={SciPost Physics},
  volume={17},
  number={3},
  pages={083},
  year={2024},
doi={ 10.21468/SciPostPhys.17.3.083}
}

@article{Gardiner1984,
  title = {{Adiabatic elimination in stochastic systems. I. Formulation of methods and application to few-variable systems}},
  author = {Gardiner, C. W.},
  journal = {Phys. Rev. A},
  volume = {29},
  issue = {5},
  pages = {2814--2822},
  numpages = {0},
  year = {1984},
  month = {May},
  publisher = {American Physical Society},
  doi = {10.1103/PhysRevA.29.2814},
  url = {https://link.aps.org/doi/10.1103/PhysRevA.29.2814}
}

@article{Ankerhold2001,
  title = {{Strong Friction Limit in Quantum Mechanics: The Quantum Smoluchowski Equation}},
  author = {Ankerhold, Joachim and Pechukas, Philip and Grabert, Hermann},
  journal = {Phys. Rev. Lett.},
  volume = {87},
  issue = {8},
  pages = {086802},
  numpages = {4},
  year = {2001},
  month = {Aug},
  publisher = {American Physical Society},
  doi = {10.1103/PhysRevLett.87.086802},
  url = {https://link.aps.org/doi/10.1103/PhysRevLett.87.086802}
}

@article{Diósi_2014,
doi = {10.1088/0031-8949/2014/T163/014004},
url = {https://doi.org/10.1088/0031-8949/2014/T163/014004},
year = {2014},
month = {dec},
publisher = {IOP Publishing},
volume = {2014},
number = {T163},
pages = {014004},
author = {Diósi, Lajos},
title = {Hybrid quantum-classical master equations},
journal = {Physica Scripta}
}

@article{Darek2017GKSL,
author = {Chru\'{s}ci\'{n}ski, Dariusz and Pascazio, Saverio},
title = {A Brief History of the {GKLS} Equation},
journal = {Open Systems \& Information Dynamics},
volume = {24},
number = {03},
pages = {1740001},
year = {2017},
doi = {10.1142/S1230161217400017},
URL = {https://doi.org/10.1142/S1230161217400017}
}

@article{Carlos2025,
doi = {10.1088/1751-8121/ae21a7},
url = {https://doi.org/10.1088/1751-8121/ae21a7},
year = {2025},
month = {nov},
publisher = {IOP Publishing},
volume = {58},
number = {48},
pages = {485303},
author = {de la Iglesia, Manuel D and Lardizabal, Carlos F},
title = {{Exact solutions of open quantum Brownian motions on the real line for two-level systems}},
journal = {Journal of Physics A: Mathematical and Theoretical}
}
\end{document}